\PassOptionsToPackage{dvipsnames}{xcolor}
\documentclass[longbibliography,nofootinbib,aps,prx,superscriptaddress,onecolumn]{revtex4-2}

\usepackage{amsmath,amsfonts,amssymb,amsthm,dcolumn,bm,qcircuit,appendix,mathtools,thmtools,thm-restate,algorithm,algpseudocode,mathrsfs,pgfplots,soul,lipsum,graphicx,braket,enumitem,caption,subcaption,ragged2e}

\usepackage{fontawesome}

\usepackage{hyperref}

\usepackage[
  labelfont=bf,
  labelsep=period,
  justification=justified,
  format=plain,
  singlelinecheck=false
]{caption}

\usepackage{pgfplots}
\usepgfplotslibrary{patchplots}
\usetikzlibrary{3d, decorations.pathmorphing}
\usepackage{tikz,tkz-euclide,tikz-3dplot}
\pgfplotsset{compat=newest}
\usepgfplotslibrary{colormaps}
\usepgfplotslibrary{patchplots}

\usetikzlibrary{matrix,fit,backgrounds,3d,arrows, decorations.pathreplacing,shapes.geometric,3d, calc}

\def\BibTeX{{\rm B\kern-.05em{\sc i\kern-.025em b}\kern-.08em
    T\kern-.1667em\lower.7ex\hbox{E}\kern-.125emX}}

\newtheorem{theorem}{Theorem}
\newtheorem{lemma}{Lemma}
\newtheorem{corollary}{Corollary}
\newtheorem{remark}{Remark}
\newtheorem{definition}{Definition}
\newtheorem{method}{Algorithm}

\numberwithin{proposition}{section}
\numberwithin{theorem}{section}
\numberwithin{lemma}{section}
\numberwithin{corollary}{section}
\numberwithin{remark}{section}
\numberwithin{definition}{section}
\numberwithin{equation}{section}

\DeclareMathOperator{\Tr}{Tr}

\newcommand{\Ibb}{\mathbb{I}}

\newcommand{\Rbb}{\mathbb{R}}

\newcommand{\norm}[1]{\left\lVert#1\right\rVert}

\usepackage{xcolor}
\usepackage{hyperref}

\hypersetup{
    colorlinks=true,
    linkcolor=MidnightBlue,
    filecolor=magenta,
    urlcolor=teal,
    pdfpagemode=FullScreen,
    citecolor=OliveGreen
}

\begin{document}
\title{Hybrid Quantum-Classical Algorithm for Hamiltonian Simulation }

\author{Nhat A. Nghiem}
\email{{nhatanh.nghiemvu@stonybrook.edu}}
\affiliation{C. N. Yang Institute for Theoretical Physics, State University of New York at Stony Brook, \\ Stony Brook, NY 11794-3840, USA}
\affiliation{Department of Physics and Astronomy, State University of New York at Stony Brook, \\ Stony Brook, NY 11794-3800, USA}

\author{Tzu-Chieh Wei}
\affiliation{C. N. Yang Institute for Theoretical Physics, State University of New York at Stony Brook, \\ Stony Brook, NY 11794-3840, USA}
\affiliation{Department of Physics and Astronomy, State University of New York at Stony Brook, \\ Stony Brook, NY 11794-3800, USA}

\begin{abstract}
We introduce a hybrid classical-quantum algorithm for simulating a Hamiltonian of the form $H= \sum_{i=1}^K H_i = \sum_{i=1}^K H_{i_1} \otimes H_{i_2} \otimes \cdots \otimes H_{i_M}$. Given that the entries of all $\{ H_{i_1}, H_{i_2} , \cdots , H_{i_M}\}$ (for all $i$) are classically known, we present a procedure (with three variants) in which these operators are classically diagonalized, and then this information is fed into three possible quantum procedures to obtain the block-encoding of $H$. The evolution operator $\exp(-iHt)$ is then obtained using the standard block-encoding/quantum singular value transformation framework. In the case where $\{H_i\}_{i=1}^K$ commute pairwise, our method can be trivially extended to the case with time-dependent coefficients. We provide a detailed discussion of the efficient regime of our hybrid framework and compare it with existing quantum simulation algorithms. Our algorithm can serve as a useful complement to existing quantum simulation algorithms, thereby expanding the reach of quantum computers for practically simulating physical systems. As a side contribution, we will show how the recent technique called \textit{randomized truncation to a quantum state} developed by Harrow, Lowe, and Witteveen  [arXiv preprint
arXiv:2510.08518, 2025] can be applied to the context of quantum simulation and particularly quantum state preparation, for which the latter can be of independent interest. 

\end{abstract}

\maketitle

\section{Introduction}
The development of quantum computation, specifically the quantum algorithm, has seen major progress in recent years. Since the early breakthroughs on the quantum factorization algorithm~\cite{shor1999polynomial}, the quantum phase estimation algorithm~\cite{kitaev1995quantum}, and the quantum search algorithm~\cite{grover1996fast}, many quantum algorithms have been proposed and developed for a wide range of tasks. Notable examples include quantum algorithms for solving linear equations~\cite{harrow2009quantum, childs2017quantum, wossnig2018quantum}, quantum machine learning algorithms~\cite{lloyd2020quantum, schuld2018supervised,schuld2019evaluating,schuld2019machine,schuld2019quantum,schuld2020circuit, biamonte2017quantum, mitarai2018quantum, havlivcek2019supervised}. Among them, simulation of quantum systems has emerged as one of the most important applications of quantum computers. The early work by Lloyd~\cite{lloyd1996universal} has shown that a quantum system can, in principle, simulate other quantum systems, i.e., quantum simulation. This work has ignited many subsequent developments~\cite{aharonov2003adiabatic, berry2007efficient, berry2012black, berry2014high,berry2015hamiltonian,berry2015simulating,berry2012black, childs2018toward, childs2019nearly, tran2020destructive, chen2021quantum}. In fact, many notable quantum algorithms, such as the quantum linear solver~\cite{harrow2009quantum, childs2017quantum} and the quantum supervised learning~\cite{lloyd2013quantum}, use quantum simulation as a core subroutine. It highlights that quantum simulation is not only an application of quantum computers but also a key primitive in the development of more advanced quantum algorithms. 


The two most popular models in the existing quantum simulation literature are the sparse-access model and the linear combination of unitary (LCU) model. In the first model, one is given oracles that can query and output information regarding the location entry in the sparse matrix, as well as its value. Some examples, in both time-independent and time-dependent settings, that fall into this model include Refs.~\cite{berry2007efficient,berry2012black, berry2014high,berry2015hamiltonian, low2017optimal,low2019hamiltonian, berry2020time, watkins2024time, an2022time, kieferova2019simulating, cao2025unifying}. The LCU model~\cite{berry2015simulating}, on the other hand, assumes that the Hamiltonian of interest is expressed as a linear combination of implementable unitaries. Within these models, multiple strategies have been proposed to obtain the evolution operator. Algorithms with optimal complexity are given in Refs.~\cite{berry2015hamiltonian, berry2015simulating, low2017optimal,low2019hamiltonian}.  

In this work, we consider the problem of Hamiltonian simulation with a different input description. The Hamiltonian of our interest has the following algebraic structure $H= \sum_{i=1}^K H_i= \sum_{i=1}^K H_{i_1} \otimes H_{i_2} \otimes \cdots \otimes H_{i_M}$ in which each $H_{i_j}$ is a small size matrix. The input description that we need for our algorithm is the \textit{classical knowledge, or the precise entry values} of the matrices $\{ H_{i_1}, ..., H_{i_M}\}$ for all $i$. We will show in detail in subsequent sections that with this classical information, we can first use a standard classical diagonalization algorithm to find the spectrum of these small matrices. Then we apply recent advances in quantum algorithms -- the block-encoding/quantum singular value transformation framework -- to block-encode these matrices and eventually block-encode the main Hamiltonian $H$ (up to some scaling), from which the evolution operator can be obtained. In addition, we will show that our hybrid algorithm can be extended in a simple manner to deal with a  time-dependent Hamiltonian of the form $  H(t) = \sum_{i=1}^K \alpha_i(t) H_i = \sum_{i=1}^K  \alpha_i(t) H_{i_1} \otimes H_{i_2} \otimes \cdots \otimes H_{i_M}$, where the Hamiltonian terms $\{H_i\}_{i=1}^K$ commute pairwise. Further, we will describe a way that makes our method efficient in the case where each $H_i$ contains many identity terms in the product of $\{H_{ij}\}_{j=1}^M$. They correspond to many physically relevant Hamiltonians, where the nontrivial part of each term in the Hamiltonian has support over a small number of sites. In particular, because the input descriptions/assumptions differ, our algorithm can complement existing quantum simulation algorithms. 

The structure of this paper is as follows. In Section~\ref{sec: setup}, we provide an overview of our work, including a description of the Hamiltonian of our interest, related assumptions, and a statement of the main results. In Section~\ref{sec: hybridalgorithm}, we outline in detail our main hybrid classical-quantum algorithm to simulate the given Hamiltonian, including the generalization to the time-dependent setting. The complexity analysis is given in Section~\ref{sec: complexityanalylsis}. Section~\ref{sec: discussion} is devoted to analyzing the most efficient regime of our work as well as the relative comparison to the existing quantum simulation literature, showing that instead of competing with them, our framework can be a well-suited complementary framework to them. The conclusion is finally given in Section~\ref{sec: conclusion}. The appendix~\ref{sec: summaryofnecessarytechniques} contains a summary and elaboration of the recipes we use in the main text.

\section{Overview: Setup, assumptions, and main results}
\label{sec: setup}
Here, we provide an overview of the problem setup, the assumptions used in our algorithms, and a brief summary of our main results. We emphasize that the benefit of our quantum algorithms for time evolution is efficiency for physically interesting Hamiltonians on the lattice, where nontrivial terms in the Hamiltonian have small support; e.g., each term in the transverse-field Ising model has support of at most two sites. Specifically, we consider a quantum system with the Hamiltonian (with a premise $||H||_o \leq 1/2$, otherwise we rescale $H$ by its maximum element) having the following form:
\begin{align}
    H = \sum_{i=1}^K H_i = \sum_{i=1}^K H_{i_1} \otimes H_{i_2} \otimes \cdots \otimes H_{i_M}.
\end{align}

Aside from the algebraic structure indicated above, there are additional assumptions that we make for our quantum algorithms.
\smallskip

\noindent {\bf Assumptions:}
\begin{itemize}
    \item For any $i \in [1,2, ..., K]$,  the classical knowledge of the Hermitian matrices $H_{i_1} , H_{i_2} , \cdots , H_{i_M} $ is provided. In other words, the classical values of the entries of these matrices are known. There are $O(KMd^2)$ such entries, where $d$ is defined next. 
    
    \item For simplicity, we assume, without loss of generality, that the dimensions of all $\{H_{i_j}\}_{j=1}^M$ are the same for all $i=1,2,...,K$. We denote such a dimension by $d$. For example, when $d=2$, the Hamiltonian above corresponds to an $M$-qubit system.
    
    \item The individual operator norm satisfies $ || H_i||_o \leq 1/2$ without loss of generality. In fact, a similar assumption is also made in most, if not all, existing quantum simulation algorithms, e.g.,~\cite{berry2007efficient,berry2012black, berry2014high,berry2015hamiltonian, childs2019nearly, low2017optimal}. 
\end{itemize}
This Hamiltonian model (with the assumptions above) captures many realistic models, for example, the transverse-field Ising model mentioned above, the Heisenberg model (any finite dimension), the toric code model,  other stabilizer models, and other nearest-neighbor lattice systems, often used in physics. 

In fact, a few quantum algorithms for simulating lattice models have been proposed, including Refs.~\cite{childs2019nearly, tran2020destructive, haah2021quantum}. However, their approaches differ from ours in many ways. 
While a more concrete summary of existing quantum algorithms shall be given in Section~\ref{sec: discussion}, we give a clarification on the key difference between our work and the prior ones. Most existing quantum simulation algorithms assume some sort of ``quantum oracle access'' to the target Hamiltonian $H$. \textcolor{black}{However, in practice, whether this oracle can be efficiently realized for any Hamiltonian is not known to us.} At the same time, we do not assume any sort of oracle access; instead, our input assumption is the \textit{classical information} about the matrices that make up $H$ as indicated above. In addition, our technique differs from existing work in the literature. For example, Ref.~\cite{berry2007efficient} breaks the initial Hamiltonian $H$ into a sum of sparse matrices, which are easy to simulate (using the oracle access). Then the simulation of $H$ can be obtained via the product formula. Ref.~\cite{berry2015simulating}, on the other hand, relies on the particular structure of $H$ (in particular, being a linear combination of unitaries) plus the truncated Taylor series. Some recent works~\cite{low2017optimal,low2019hamiltonian}, which lay the foundation of QSVT, again rely on the assumption that there is oracle access to the entries of $H$. In summary, most existing works either assume the oracle access to the Hamiltonian or consider Hamiltonians that are assumed to have a known linear combination of unitaries. In contrast, we make use of the classical knowledge to classically diagonalize $\{ H_{i_1}, H_{i_2}, \cdots, H_{i_M} \} $ for all $i$. This information is then fed to a quantum procedure to block-encode all these matrices, and eventually the total Hamiltonian $H$, from which the evolution operator $\exp(- i H t)$ can be obtained using a standard recipe from the block-encoding/QSVT framework. For convenience, we summarize our main results in the following theorem.
\begin{theorem}
\label{thm: mainresult}
    Let $H = \sum_{i=1}^K H_i = \sum_{i=1}^K H_{i_1} \otimes H_{i_2} \otimes \cdots \otimes H_{i_M}$ with the assumptions given above. For an $i$, let $R_i$ be the set of nontrivial (i.e., non-identity) matrices among $ H_{i_1} , H_{i_2} , \cdots , H_{i_M} $, and accordingly $|R_i|$ be the size of this set. Define $|R| \equiv \{ |R_i| \}_{i=1}^K$. Then the evolution operator $\exp\big( - i H t  \big)$ can be obtained up to an additive precision $\delta$ by a hybrid classical-quantum algorithm with the complexity for each part as follows:
    \begin{itemize}
        \item (First approach)[Sec.~\ref{sec: quantum}] Classical time complexity $\mathcal{O}\Big( |R| K  d^3 \Big) $ --- Quantum circuit complexity 
        $$\mathcal{O}\Big(  K^2 d^{|R|}  M \log(d)    t \log \frac{1}{\delta}\Big).   $$
        \item (Second approach) [Sec.~\ref{sec: alternativeapproach}] Classical time complexity $\mathcal{O} \Big( |R| K d^3 \Big)$ --- Quantum circuit complexity 
        $$ \mathcal{O}\Big(   K^2 |R| \log(d)   \frac{ d^{4|R|}}{\delta^2} t^3 \log \frac{1}{\delta} \Big). $$
        \item (Third approach -- applicable for nonnegative $\{H_i \}_{i=1}^K$) [Sec.~\ref{sec: thirdapproach}] Classical complexity $\mathcal{O}\Big(   K^2 s^{|R|}  |R| d^{|R|} \rm poly(d) \Big)$ --- Quantum circuit complexity  
        $$\mathcal{O}\Big( t  K^2 d^{|R|} |R| \log (Lsd) \log \frac{1}{\delta}  \Big), $$
        where $s \ll d$ and can be chosen.
    \end{itemize}
\end{theorem}
In particular, our algorithms can be easily extended to the time-dependent setting where $\{H_i\}_{i=1}^K$ commute.
\begin{corollary}
    Let $H = \sum_{i=1}^K \alpha_i(t) H_i = \sum_{i=1}^K \alpha_i(t) H_{i_1} \otimes H_{i_2} \otimes \cdots \otimes H_{i_M}$ with the assumptions given above, plus another assumption that all pairs of $\{H_i\}_{i=1}^K$ commute and each $\alpha_i(t)$ is an efficiently integrable function of $t$. Then the evolution operator can be obtained by a hybrid algorithm with the following classical and quantum complexity:
    \begin{itemize}
        \item Classical complexity  $ \mathcal{O}\Big( MKd^3 \Big)  $ --- Quantum circuit complexity $ \mathcal{O}\Big( K^2 d^{2|R|}  t |R| \log(d)   \log \frac{1}{\delta}\Big) $.
    \end{itemize}
\end{corollary}

A more detailed analysis and discussion of our algorithm will be given in Sect.~\ref{sec: discussion}. Here, we point out that our hybrid algorithm achieves the best performance when $R = \mathcal{O}(1)$, which is possible in practice, as many Hamiltonians from quantum many-body contexts satisfy this condition; e.g., lattice systems with nearest-neighbor interactions fall into this case. In particular, in the same Sec.~\ref {sec: discussion}, we will discuss that our hybrid framework can be complementary to the existing quantum simulation algorithms, including general time-independent settings, time-dependent settings, and lattice simulations. 

In addition to these, our work also contains two byproducts. First, we provide a concrete scenario in which the recent result of Ref.~\cite{harrow2025randomized} can be applied to the context of quantum simulation, thus providing a positive answer to one of the open aspects mentioned in their work. Second, we particularly show how the same result as in Ref.~\cite{harrow2025randomized} can be used to practically enhance the state preparation protocol as proposed in Ref.~\cite{zhang2022quantum}. Generally speaking, to prepare a $s$-sparse, $n$-qubits quantum state $\ket{\Phi} \in \mathbb{C}^{2^n}$, the method of~\cite{zhang2022quantum} is universal (which means that it can prepare any known state without any specific structure), achieving a quantum circuit depth $\mathcal{O}(\log sn)$, using extra $\mathcal{O}(s)$ ancilla qubits. As such, this method is only practical when the sparsity $s$ is sufficiently small, e.g., being $\rm poly(n)$. When the sparsity is high or the state $\ket{\Phi}$ is dense, then this method becomes costly. We will show in greater detail in Sec.~\ref{sec: discussion} how to leverage the result of~\cite{harrow2025randomized} to make the method of~\cite{zhang2022quantum} more accessible to dense states, achieving a quantum circuit of highly efficient depth plus an efficient ancilla usage, at the cost of employing more classical resources and incurring some error tolerance. Here, we summarize it in the following corollary:

\begin{corollary}
\label{corollary: statepreparation}
 Let $\ket{\Phi} \in \mathbb{C}^{2^n}$ be an $n$-qubits quantum state of interest. Fix $S$ to be some desired sparsity parameter (with $S \ll 2^n$ can be chosen) and a desired error tolerance $\epsilon$. Then there is a procedure involving a classical algorithm with complexity at most $2^n$ and a quantum circuit of depth $\mathcal{O}\Big(L \log (Sn) \Big)$ plus $\mathcal{O}(S )$ ancilla qubits, that can prepare the state $\ket{\tilde{\Phi}}$ s.t. $\big|\big| \ket{\tilde{\Phi}} - \ket{\Phi}\big|\big|_2 \leq \epsilon $. The value of $L$ depends on the entries of $\ket{\Phi}$ and, in general, it does not explicitly depend on $n$. 
\end{corollary}

\section{Hybrid algorithm for simulating $H$}
\label{sec: hybridalgorithm}

As mentioned earlier, our algorithm is hybrid, involving both classical and quantum procedures. Roughly speaking, by first leveraging the classical knowledge of $H_{i_1} , H_{i_2} , \cdots , H_{i_M}  $, we can use any existing classical algorithms to diagonalize them and find the corresponding eigenvalues and eigenvectors, for each $H_{i_k}$ separately. Then, by exploiting the tensor structure in each $H_i=H_{i_1} \otimes H_{i_2} \otimes \cdots \otimes H_{i_M}$, we can build the entire spectrum of $H_i$. The final step is to leverage the classical information (of the spectrum of $\{H_{i_j}\}_{j=1}^M$) combined with the recent advances in quantum algorithmic frameworks, e.g., block-encoding and quantum singular value transformation (QSVT), to obtain the (block-encoding of) $\{H_i\}$, then of $H$, then the desired evolution operator $\exp(-i H t)$. We have provided a summary of the necessary recipes from the block-encoding and QSVT in Appendix~\ref{sec: summaryofnecessarytechniques}. 

\subsection{Classical component}
\label{sec: classical}
As mentioned above, the first step is to use existing classical algorithms to find the spectrum and eigenstates of $H_{i_j}$ for $j=1,2,...,M$. We denote the eigenvalues/eigenvectors of $H_{i_j}$ as $ \{ \lambda_{i_j}(k) / \ket{\lambda_{i_j}(k)} \}_{k=1}^d $.  As $H_{i_j}$ is of dimension $d \times d$, the maximum rank of this matrix is $d$. If the rank of $H_{i_j}$ is lower than $d$, then we understand that for all those $k$ between $r(H_{i_j}) \equiv \rm rank\  H_{i_j}$ and $d$, the eigenvaluse vanish, i.e., $ \lambda_{i_j}(k) = 0$.   

As $H_i = \bigotimes_{j=1}^M H_{i_j}$, a pair of eigenvalues and eigenvectors of $H_i$ is, respectively:
\begin{align}
   \lambda_{i_1}(k_1) \lambda_{i_2}(k_2) \cdots \lambda_{i_M} (k_M) , \\
   \ket{\lambda_{i_1}(k_1)}\otimes \ket{ \lambda_{i_2}(k_2)} \cdots \ket{\lambda_{i_M} (k_M)},
\end{align}
for $k_1,k_2,...k_M \in [1,2,...,d]$. As such, the matrix $H_i$ can be written in the spectral decomposition as follows,
\begin{align}
    H_i = \sum_{k_1,k_2,...,k_M = 1}^d \big(\lambda_{i_1}(k_1) \lambda_{i_2}(k_2) \cdots \lambda_{i_M} (k_M) \big)  \ket{\lambda_{i_1}(k_1)} \bra{ \lambda_{i_1}(k_1)}\otimes \ket{ \lambda_{i_2}(k_2)}  \bra{ \lambda_{i_2}(k_2)} \cdots \otimes\ket{\lambda_{i_M} (k_M)} \bra{ \lambda_{i_M} (k_M)}.
\end{align}
The above sum, in principle, includes both nonzero and zero eigenvalues $  \{ \lambda_{i_j}(k)  \}_{k=1}^d$ for $j=1,2,...,M$. But, in practice, only those with nonzero eigenvalues are relevant. Hence, effectively, the total of terms in the above summation is $r(H_{i_1}) r(H_{i_2}) \cdots r(H_{i_M})$, which is exactly $r(H_i)$. 

\subsection{Quantum component: approach 1}
\label{sec: quantum}
In the previous section, we have used the classical algorithm to find the spectrum and eigenstates of $\{H_{i_j}\}_{j=1}^M$ and from them we can re-construct  $H_i$ from the spectral decomposition. As the first step of our quantum algorithm, we need to prepare the state:
\begin{align}
    \ket{\lambda_{1}(k_1)}\otimes \ket{ \lambda_{2}(k_2)} \cdots \otimes \ket{\lambda_{M} (k_M)}.
\end{align}
Then the next step is to prepare the (block-encoding of) the density operator:
\begin{align}
    \ket{\lambda_{1}(k_1)} \bra{ \lambda_{1}(k_1)}\otimes \ket{ \lambda_{2}(k_2)}  \bra{ \lambda_{2}(k_2)} \cdots \ket{\lambda_{M} (k_M)} \bra{ \lambda_{M} (k_M)},
\end{align}
which will allow us to do linear combination.

To this end, we mention a few necessary recipes. The first one is regarding quantum state preparation:
\begin{lemma}[State preparation \cite{zhang2022quantum}]
\label{lemma: statepreparation}
    Let $\ket{\Phi}$ be an $N$-dimensional state with classically known entries and sparsity $s$ (the number of nonzero entries). Then $\ket{\Phi}$ can be prepared using a quantum circuit of depth $\mathcal{O}\left( \log (s \log N) \right)$, plus extra $\mathcal{O}(s\log(s)\log N)$ ancilla qubits. 
\end{lemma}
We note that there is a rich literature on quantum state preparation~\cite{zhang2022quantum, nakaji2022approximate, zoufal2019quantum, grover2002creating, mcardle2022quantum, zhang2021low, plesch2011quantum, marin2023quantum}, each work possessing a certain strength and limitation. For example, the lemma above, which is based on~\cite{zhang2022quantum}, provides a universal approach to prepare a known state. It means that this method can be applied to a state of arbitrary structure, as long as the entries are classically provided. However, the circuit complexity, including both depth and width, is only efficient when $s$ is not too large, e.g., $\in \mathcal{O}(\log N)$. On the other hand, the work of Ref.~\cite{nakaji2022approximate} introduces a variational approach for state preparation. Generally, their approach requires a circuit of gate complexity $\mathcal{O}\left( \log N \right)$ (we are ignoring some other factors), and in principle can deal with non-sparse states. However, it comes with a heuristic performance and an error tolerance. Ref.~\cite{grover2002creating} also proposes an $\mathcal{O}(\log N)$ complexity circuit to prepare a state, provided that the amplitudes of such a state obey certain distribution functions. This directly implies that the method is effective only for certain states, which is apparently a limitation.  The Ref.~\cite{mcardle2022quantum} introduces a QSVT-based approach for state preparation. This method can also deal with the non-sparse states in general, requiring a quantum circuit of complexity $\mathcal{O}( \log N)$, plus $\mathcal{O}(1)$ ancilla qubit. However, the method of~\cite{mcardle2022quantum} only works when the amplitudes of the desired state are output of a smooth and efficiently computable function. A similar performance can also be found in Ref.~\cite{marin2023quantum}, in which a quantum circuit of complexity $\mathcal{O}(\log N)$ is shown to be able to prepare a state, given that the amplitudes form an efficiently integrable function.  

The second recipe we need is the following result regarding the block-encoding of a density operator:

\begin{lemma}[Block encoding of density matrix, see e.g.~\cite{gilyen2019quantum}]
\label{lemma: improveddme}
Let $\rho = \Tr_A \ket{\Phi}\bra{\Phi}$, where $\rho \in \mathbb{H}_B$ and $\ket{\Phi} \in  \mathbb{H}_A \otimes \mathbb{H}_B$. Given a unitary $U$ that prepares $\ket{\Phi}$ from $\ket{\bf 0}_A \otimes \ket{\bf 0}_B$, there exists a highly efficient procedure that constructs an exact unitary block encoding of $\rho$ using $U$ and $U^\dagger$ once each plus $\mathcal{O}\big( \log \dim \rho \big)$ another 2-qubit gates. 
\end{lemma}

With the above tools, we are now ready to describe in detail the quantum algorithm to simulate $H$, provided that the classical knowledge of the spectrum of $\{ H_{i_j} \}_{j=1}^M$  is obtained via the classical algorithm. For subsequent usage, we define $\gamma_i = \sum_{\substack{k_1,k_2,...,k_M = 1 }}^d |\lambda_{i_1}(k_1) \lambda_{i_2}(k_2) \cdots \lambda_{i_M} (k_M) |$. We note that this summation also includes the zero eigenvalues, which implies that $\gamma_i = 0$ if any of the eigenvalues $\{\lambda_{i_j}\}_{j=1}^M$ being zero.  
\begin{method}
\label{algo: timeindependentsimulation}
    The algorithm for simulating time-independent $H=\sum_{i=1}^K H_{i_1} \otimes H_{i_2} \otimes \cdots \otimes H_{i_M} $ proceeds as follows:
\begin{enumerate}
    \item Use Lemma~\ref{lemma: statepreparation} to obtain, say $U_{k_1},U_{k_2},...,U_{k_M}$ that prepares the state $\ket{\lambda_{i_1}(k_1)},  \ket{ \lambda_{i_2}(k_2)} \cdots , \ket{\lambda_{i_M} (k_M)} $. Then it is clear that $ U_{k_1} \otimes U_{k_2} \otimes \cdots \otimes U_{k_M}$ prepares the state $\ket{\lambda_{i_1}(k_1)}\otimes \ket{ \lambda_{i_2}(k_2)} \cdots \otimes \ket{\lambda_{i_M} (k_M)} $. We note that in the case where $d = \mathcal{O}(1)$  (e.g., $d=2,3$, then each composed system corresponds to a qubit, or qudit system), each $\ket{\lambda_{i_j} (k_j)}$, in principle, a single $U(d)$ rotation gate is sufficient to prepare such a state without ancilla. As such, we do not need to use Lemma \ref{lemma: statepreparation}, which requires some ancilla qubits. 

    \item 
    Using the above lemma (\ref{lemma: improveddme})  with this unitary  $U_{k_1} \otimes U_{k_2} \otimes \cdots \otimes U_{k_M}$ allows us to obtain an exact block-encoding of 
\begin{align}
    \ket{\lambda_{i_1}(k_1)}\otimes \ket{ \lambda_{i_2}(k_2)} \cdots \otimes \ket{\lambda_{i_M} (k_M)} \ \bra{\lambda_{i_1}(k_1)}\otimes \bra{ \lambda_{i_2}(k_2)} \cdots \otimes \bra{\lambda_{i_M} (k_M)}.
\end{align}

Repeat the same procedure for different $k_1,k_2,...,k_M$, we then obtain the block-encoding of
$$  \{ \ket{\lambda_{i_1}(k_1)}\otimes \ket{ \lambda_{i_2}(k_2)} \cdots \otimes \ket{\lambda_{i_M} (k_M)} \ \bra{\lambda_{i_1}(k_1)}\otimes \bra{ \lambda_{i_2}(k_2)} \cdots \otimes \bra{\lambda_{i_M} (k_M)}\} _{\substack{k_1,k_2,...,k_M = 1 }}^d. $$
The next step is to use Lemma \ref{lemma: sumencoding} with combination factors $\{  \frac{ \lambda_{i_1}(k_1) \lambda_{i_2}(k_2) \cdots \lambda_{i_M} (k_M)}{\gamma_i} \}_{\substack{k_1,k_2,...,k_M = 1 }}^d $ to construct the block-encoding of:
\begin{align}
        &\!\!\!\!\!\!\!\!\!\!\!\!\!\!\!\!\sum_{\substack{k_1,k_2,...,k_M = 1 }}^d\frac{ \lambda_{i_1}(k_1) \lambda_{i_2}(k_2) \cdots \lambda_{i_M} (k_M)}{\gamma_i} \Big( \ket{\lambda_{i_1}(k_1)}\otimes \ket{ \lambda_{i_2}(k_2)} \cdots \otimes \ket{\lambda_{i_M} (k_M)} \Big) \Big(\bra{\lambda_{i_1}(k_1)}\otimes \bra{ \lambda_{i_2}(k_2)} \cdots \otimes \bra{\lambda_{i_M} (k_M)} \Big) \nonumber\\
        &= \frac{1}{\gamma_i}  H_{i_1} \otimes H_{i_2} \otimes \cdots \otimes H_{i_M},
    \end{align}
    \label{eqn: 11}
    which is exactly $\frac{H_i}{\gamma_i}$. 
    
    \item In a similar manner, repeat the above steps for $i=1,2,...,K$. Then we obtain an $\epsilon$-approximated block-encoding of $\{ \frac{H_i}{\gamma_i}\}_{i=1}^K$. 
    \item Use Lemma \ref{lemma: sumencoding} again with the linear combination factors to be $\{ \frac{\gamma_i}{\sum_{i=1}^K \gamma_i} \}_{i=1}^K$ and the block-encodings of $\{ \frac{H_i}{\gamma_i}\}_{i=1}^K $ to construct the $\epsilon$-approximated block-encoding of: 
    \begin{align}
        \sum_{i=1}^K \frac{\gamma_i}{ \sum_{i=1}^K\gamma_i} \frac{H_i}{\gamma_i} = \frac{1}{\sum_{i=1}^K\gamma_i} H.
    \end{align}
    As a brief comment, the block-encoded operator above admits an approximation error of $\epsilon$ instead of $K\epsilon$. At first glance, due to the linear accumulation from $K$ block-encoded operators $\{ \frac{H_i}{\gamma_i}\}_{i=1}^K$, each with a $\epsilon$-approximation, the resulting operator should have had an accumulated error $K\epsilon$. However, the linear combination comes with the factors $\{ \frac{\gamma_i}{\sum_{i=1}^K \gamma_i} \}_{i=1}^K $, so the totally accumulated error is: 
    \begin{align}
        \sum_{i=1}^K \frac{\gamma_i}{\sum_{i=1}^K \gamma_i} \epsilon = \epsilon.
    \end{align} 
    \item Use Lemma \ref{lemma: amp_amp} with the block-encoding above to remove the factor $ \sum_{i=1}^K \gamma_i$, resulting in the block-encoding of $H$. 
    
    Before proceeding to the final step, we invoke the following Lemma:
    \begin{lemma}\label{lemma: qsvt}[\cite{gilyen2019quantum} Theorem 56]
\label{lemma: theorem56}
Suppose that $U$ is an
$(\alpha, a, \epsilon)$-encoding of a Hermitian matrix $A$. (See Definition 43 of~\cite{gilyen2019quantum} for the definition.)
If $P \in \mathbb{R}[x]$ is a degree-$p$ polynomial satisfying that
\begin{itemize}
\item for all $x \in [-1,1]$: $|P(x)| \leq \frac{1}{2}$.
\end{itemize}
Then, there is a quantum circuit $\tilde{U}$, which is an $(1,a+2,4p \sqrt{\frac{\epsilon}{\alpha}})$-encoding of $P(A/\alpha)$, and consists of $p$ applications of $U$ and $U^\dagger$ gates, a single application of controlled-$U$ and $\mathcal{O}((a+1)d)$
other one- and two-qubit gates.
\end{lemma}
\item The final step is to use the above lemma with the polynomial $P$ being the Jacobi-Anger expansion (as used in the original quantum signal processing/QSVT work~\cite{low2017optimal,low2019hamiltonian}) that approximates the function $\exp(-i xt)$ in the domain $x \in [-1,1]$. It allows us to transform the block-encoding of $H$ to the block-encoding of $\exp\Big(  - i H t  \Big)$. 
\end{enumerate} 
\end{method}
The algorithm above is generic in the sense that it works for any structure of each $H_i$. Below, we will show how this algorithm can be simplified in the scenario where the tensor decomposition of $H_i$ contains many terms that can be easily block-encoded, for example, being the identity matrix. We shall show that in such a case we only need to run the algorithm above with those nontrivial terms, and thus save a large amount of resources, as will be discussed in detail in Section~\ref{sec: complexityanalylsis}.

\subsection{Simplification in special cases}
\label{sec: simplification}

 In the above algorithms, a crucial component is to use the classical algorithm to diagonalize the matrices $\{ H_{i_1}, H_{i_2}, ..., H_{i_M}\}$ for all $i$. Then this classical information is then fed into a quantum procedure (e.g., combined with Lemma~\ref{lemma: improveddme}) to obtain the block-encoding of these matrices, then the block-encoding of $H_i$ (up to a factor), and finally the block-encoding of $H$. Here we point out that, in certain cases, these steps can be simplified, and thus the complexity is accordingly improved:
\begin{itemize}
    \item For any $i$, when some of $H_{i_1}, H_{i_2}, ..., H_{i_M}$ admit a simple structure so that their spectrum, including eigenvalues and eigenvectors, can be easily obtained without doing exact diagonalization. In this case, we do not need to run the classical diagonalization algorithm for them, thus saving a certain classical computational time. An example can be when some $H_{i_j} = a \ket{0}\bra{0} + b\ket{1}\bra{1}$, which has eigenvector $\ket{0}, \ket{1}$ with eigenvalue $a,b$, respectively.
    \item For any $i$, when some of $H_{i_1}, H_{i_2}, ..., H_{i_M}$ can be easily, or more broadly, can be efficiently block-encoded. The most trivial example is when they are identity operators, which can be block-encoded by a circuit of $\mathcal{O}(1)$  complexity (see Def.~\ref{def: blockencode} and the discussion below). As a result, the complexity for the quantum part can be further optimized because we do not need to diagonalize these trivial terms and run the block-encoding procedure. Instead, we only need to care about those nontrivial terms. 
\end{itemize}
So, in the second case, where $H_i = \otimes_{j=1}^M H_{i_j}$ and some of these composing terms are assumed to be trivial for simplicity, the Algorithm~\ref{algo: timeindependentsimulation} can be simplified as follows. For a given $i$, let $R_i$ denote the set of nontrivial terms among $H_{i_1}, ..., H_{i_M}$. We consider the following Hamiltonian instead: $H_i' = \otimes_{H_{i_j} \in R_i} H_{i_j}$. Then, by running the first 2 steps of the Algorithm~\ref{algo: timeindependentsimulation}, we obtain the block-encoding of $\frac{H_i'}{\gamma_i'}$ (where $\gamma_i'$ now is defined as the sum of absolute eigenvalues of those operators appearing in $H_i'$ only). Then, via Lemma~\ref{lemma: tensorproduct} plus the fact that identity matrices (of arbitrary dimension) can be block-encoded, we can build the block encoding of 
\begin{align}
    \frac{H_i'}{\gamma_i'} \otimes \Ibb  = \frac{1}{\gamma_i'} \big(\otimes_{H_{i_j} \in R_i} H_{i_j}\big)  \otimes  \underbrace{\Ibb_d \otimes \Ibb_d \otimes \cdots \Ibb_d}_{M-|R_i| \text{\ terms}}.
\end{align}
From this block encoding, we can use it with Lemma~\ref{lemma: product} to multiply it with the SWAP operators (which block-encodes itself as it is a unitary), which appropriately swap the location of $H_{i_j}$'s (for ${H_{i_j} \in R_i}  $) with $\Ibb_d$, so that the resulting operator is exactly $H_i$. As a simple illustrative example, we consider $H_i = H_{i_1} \otimes \Ibb \otimes H_{i_2} \otimes \Ibb$. First, we execute the two steps of the algorithm above to build the block-encoding of $\varpropto H_{i_1} \otimes H_{i_2}$ (ignoring the factor $\gamma_i$ in the Algo.~\ref{algo: timeindependentsimulation} for simplicity). Then we build the block-encoding of $ \varpropto   H_{i_1} \otimes H_{i_2} \otimes \Ibb_d \otimes \Ibb_d$, followed by multiplying it with the SWAP operator (between the second and third systems) to build the block-encoding of $ \varpropto   H_{i_1} \otimes \Ibb_d \otimes H_{i_2} \otimes \Ibb_d $. It can be seen that the number of SWAP operations is upper bounded by the number of nontrivial terms $|R_i|$.

We note that in this case, if we do not use the above procedure to simplify, instead, we run the Algorithm \ref{algo: timeindependentsimulation}, then $\gamma_i$ is related to $\gamma_i'$ 
$$ \gamma_i = d^{M- |R_i|} \gamma_i'. $$
By using this relation, there is a substantial saving in the computational procedure. 

\subsection{A second approach to prepare the block-encoding of $\frac{H_i}{\gamma_i}$   }
\label{sec: alternativeapproach}
In this section, we outline another approach to prepare the block-encoding of $\rho \equiv \frac{H_i}{\gamma_i}$ (step 2 of the algorithm above). The general idea is as follows. Consider a classical probability distribution over $M$ discrete variables 
\begin{align}
    p_i(k_1,k_2,...,k_M) = \frac{ |\lambda_{i_1}(k_1) \lambda_{i_2}(k_2) \cdots \lambda_{i_M} (k_M) |}{\gamma_i}.
\end{align}
As we noted earlier, this value is only nonzero if all eigenvalues $\{ \lambda_{i_1}(k_1), ..., \lambda_{i_M}(k_M) \}$ are nonzero and we note the eigenvalues can be negative, so in the above, we need to take the absolute value to make sure that the probability is nonnegative. Suppose that we opt to prepare the state $ \ket{\lambda_{i_1}(k_1)}\otimes \ket{ \lambda_{i_2}(k_2)} \cdots \otimes \ket{\lambda_{i_M} (k_M)}$ with the probability $ p_i(k_1,k_2,...,k_M) $. Then we use Lemma \ref{lemma: improveddme} to block-encode 
$$ \ket{\lambda_{i_1}(k_1)}\otimes \ket{ \lambda_{i_2}(k_2)} \cdots \otimes \ket{\lambda_{i_M} (k_M)} \bra{\lambda_{i_1}(k_1)}\otimes \bra{ \lambda_{i_2}(k_2)} \cdots \otimes \bra{\lambda_{i_M} (k_M)},$$
which can be easily turned into the block-encoding of 
$$ -\ket{\lambda_{i_1}(k_1)}\otimes \ket{ \lambda_{i_2}(k_2)} \cdots \otimes \ket{\lambda_{i_M} (k_M)} \bra{\lambda_{i_1}(k_1)}\otimes \bra{ \lambda_{i_2}(k_2)} \cdots \otimes \bra{\lambda_{i_M} (k_M)}.$$
We consider the procedure which selects and obtains the block-encoded operator
$$ \pm \ket{\lambda_{i_1}(k_1)}\otimes \ket{ \lambda_{i_2}(k_2)} \cdots \otimes \ket{\lambda_{i_M} (k_M)} \bra{\lambda_{i_1}(k_1)}\otimes \bra{ \lambda_{i_2}(k_2)} \cdots \otimes \bra{\lambda_{i_M} (k_M)}$$
with probability $ p_i(k_1,k_2,...,k_M)$, with the $+$ sign corresponds to positive $\lambda_{i_1}(k_1) \lambda_{i_2}(k_2) \cdots \lambda_{i_M} (k_M) $, and $-$ sign otherwise. Then this procedure produces the following block-encoded operators on average 
$$ \rho =\sum_{\substack{k_1,k_2,...,k_M = 1 }}^d  \pm p_i(k_1,k_2,...,k_M)  \Big( \ket{\lambda_{i_1}(k_1)}\otimes \ket{ \lambda_{i_2}(k_2)} \cdots \otimes \ket{\lambda_{i_M} (k_M)} \Big) \Big(\bra{\lambda_{i_1}(k_1)}\otimes \bra{ \lambda_{i_2}(k_2)} \cdots \otimes \bra{\lambda_{i_M} (k_M)} \Big), $$
where the sign $\pm$ depends on whether the corresponding eigenvalue $\lambda_{i_1}(k_1), ..., \lambda_{i_M}(k_M) $ is positive or negative. It can be seen that the above operator is exactly $\frac{H_i}{\gamma_i}$. 
By repeating the procedure above $N$ times, suppose that at the $j$-th step we obtain the block-encoding of 
$$ \rho_j =\pm  \Big( \ket{\lambda_{i_1}(k_1)}\otimes \ket{ \lambda_{i_2}(k_2)} \cdots \otimes \ket{\lambda_{i_M} (k_M)} \Big)_j \Big(\bra{\lambda_{i_1}(k_1)}\otimes \bra{ \lambda_{i_2}(k_2)} \cdots \otimes \bra{\lambda_{i_M} (k_M)} \Big)_j,$$
then we can use Lemma~\ref{lemma: sumencoding} to construct the block-encoding of $ \sum_{j=1}^N \frac{\rho_j}{N}$. Our next step is to show that for a suitable choice of $N$, this operator can approximate $\rho$.

To this end, we recall the following well-known results, namely, the central limit theorem:
\begin{lemma}
    Let $\{X_i\}_{i=1}^{\infty}$ be a sequence of \textit{independent and identically distributed (i.i.d.) real-valued random variables} such that
    \begin{align}
        \mu = \mathbb{E}[X] , \ \sigma =  \rm Var (X) > 0 
    \end{align}
    are both finite and existing. Define the partial sums $S_N = \sum_{i=1}^N X_i$. Then the normalized sum $Z_N = \frac{S_N- N\mu}{\sigma \sqrt{N}}$ converges in distribution to a standard normal random  variable, i.e., for every $x\in \Rbb$:
    \begin{align}
        \lim_{n \longrightarrow \infty} \mathbb{P} \Big(  \frac{S_N - N \mu}{\sigma \sqrt{N}}  \leq x \Big) = \Phi (x),
    \end{align}
    where $\Phi(x) = \frac{1}{2\pi} \int_{-\infty}^x e^{-t^2/2} dt$ is the cumulative distribution function of the standard normal distribution. 
\end{lemma}
A more friendly lower bound statement of the central limit theorem, which is valid for all cases, is the Chebyshev bound:
\begin{align}
    \mathbb{P} \Big(  \Big| \frac{S_N}{N  } - \mu \Big| \leq \epsilon \Big)  \geq 1- \frac{\sigma^2}{N \epsilon^2}. 
\end{align}
So, by choosing $N = \mathcal{O}(\frac{1}{\epsilon^2})$, it can be guaranteed that the probability of having the deviation $ \Big| \frac{S_N}{N  } - \mu \Big|$ less than $\epsilon$ is $\mathcal{O}(1)$. 

To apply the above result to our case, we observe that for $N = \mathcal{O}( \frac{1}{\epsilon^2})$, 
\begin{align}
\mathbb{P}\Big( \Big|  \sum_{j=1}^N \frac{\rho_j}{N} - \rho    \Big|_{\infty} \leq \epsilon \Big) \geq 1- \frac{\sigma^2}{N \epsilon^2}, 
\end{align}
where $|.|_{\infty}$ refers to the maximum entry value in magnitude. We would like to elaborate on the above inequality that, in this case, we treat each entry of $\rho$ as a random variable, and use the central limit theorem for each entry. Then all the deviations among all entries are bounded by $\epsilon$, which implies that the maximum of them is also bounded by $\epsilon$. 

It is a known property~\cite{golub2013matrix} between the operator/spectral norm (maximum singular/eigenvalue in magnitude) and the infinity norm above that, for a squared matrix $D$ having sparsity $s_D$ (the maximum number of nonzero entries in each row/column), it holds that:
\begin{align}
    \big|D  \big|_{o}  \leq \big| D \big|_{\infty} s_D.
\end{align}
In our case, the inequality above can be used to show that:
\begin{align}
    \Big|  \sum_{j=1}^N \frac{\rho_j}{N} - \rho    \Big|_{o} \leq \Big|  \sum_{j=1}^N \frac{\rho_j}{N} - \rho    \Big|_{\infty}  s_\rho,
    \label{eqn: 18}
\end{align}
where $s_\rho$ is defined as the sparsity of $\sum_{j=1}^N \frac{\rho_j}{N} - \rho $. We would like to note that, in principle, $s_\rho$ can be large, and generally computing $s_\rho$ is not efficient as we would need to perform the tensor products within $\rho_j$ for all $j$, followed by a matrix subtraction by $\rho$. The complexity of this procedure can grow as much as $d^M$, unless all of the eigenvectors of $\rho_j$ are very sparse, which seems unlikely. Below, we  discuss a way to bound the sparsity $s_\rho$, and how it can be reduced in certain cases, in light of the discussion of Section~\ref{sec: simplification}.  \\

\noindent
\textbf{Bounding the sparsity $s_\rho$.} We remind the reader that for a given $i$, the value of $s_\rho$ is defined as the sparsity $\sum_{j=1}^N \frac{\rho_j}{N} - \rho $. In general, since we know (via classical diagonalization) the sparsity of $\ket{\lambda_{i_1}(k_1)} , \ket{ \lambda_{i_2}(k_2)} \cdots , \ket{\lambda_{i_M} (k_M)} $, then the sparsity $s(\rho_j)$ of 
$$ \rho_j = \Big( \ket{\lambda_{i_1}(k_1)}\otimes \ket{ \lambda_{i_2}(k_2)} \cdots \otimes \ket{\lambda_{i_M} (k_M)} \Big)_j \Big(\bra{\lambda_{i_1}(k_1)}\otimes \bra{ \lambda_{i_2}(k_2)} \cdots \otimes \bra{\lambda_{i_M} (k_M)} \Big)_j$$
can be computed accordingly by multiplying the sparsity of the composing matrices. In addition, we know the sparsity of $\rho = H_i/\gamma_i$, which can be computed by multiplying the sparsity of $H_{i_1},..., H_{i_M}$). The value of $s_\rho$ can be upper-bounded as 
$$s_\rho \leq \sum_{j=1}^N s\big( \frac{\rho_j}{N} \big) + s(\rho),$$ 
which can be seen as follows. Suppose $A,B$ are two matrices of the same dimension. In the worst case, where the location of all entries of $A$ and $B$ are different (e.g., for $A_{ij}  \neq 0$ then $B_{ij} = 0$), then $A \pm B$ has the sparsity equal to the sparsity of $A$ plus the sparsity of $B$. Using similar reasoning, it can be shown that the sparsity of sum of matrices is upper bounded by the sum of sparsity of these matrices.

Whether the value of $s_\rho$ is high or not depends on the sparsity of each $\rho_j$, and thus depends on the sparsity of those eigenstates $\ket{\lambda_{i_1}(k_1)} , \ket{ \lambda_{i_2}(k_2)} \cdots , \ket{\lambda_{i_M} (k_M)} $. In the worst case, if the sparsity of these vectors is $\mathcal{O}(d)$, then the sparsity $s(\rho_j)$ of each $\rho_j$ can be as high as $\mathcal{O}(d^M)$. 

\smallskip
\noindent
\textbf{Simplification in special cases (as in Section \ref{sec: simplification}). } We note that the algorithm given in this section is still assuming that all matrices $H_{i_1}, H_{i_2}, ..., H_{i_M}$ are nontrivial. In Section~\ref{sec: simplification}, we discussed that when some of these matrices are trivial, the algorithm in Section~\ref{sec: quantum} can be simplified, since we only need to consider the nontrivial terms. As such, it is reasonable to expect that the method outlined in this section can also gain benefit in those cases. Indeed, we can do exactly the same thing as we described in Section~\ref{sec: simplification}. First, we redefine the Hamiltonian $H_i' = \otimes_{H_{i_j} \in R_i} H_{i_j}$. Then we execute the same procedure as we discussed in Section \ref{sec: simplification}, to obtain the block-encoding of
$$ \rho_j' = \pm \Big( \ket{\lambda_{i_1}(k_1)}\otimes \ket{ \lambda_{i_2}(k_2)} \cdots \otimes \ket{\lambda_{i_{|R_i|}} (k_{|R_i|})} \Big)_j \Big(\bra{\lambda_{i_1}(k_1)}\otimes \bra{ \lambda_{i_2}(k_2)} \cdots \otimes \bra{\lambda_{i_{|R_i|}} (k_{|R_i|})} \Big)_j.$$
As analyzed before, the approximation of $\rho = \frac{H_i'}{\gamma_i'}$ can be taken as $\sum_{j=1}^N \frac{\rho_j}{N}$ and thus the (approximated) block-encoding of $\rho$ can be obtained via Lemma~\ref{lemma: sumencoding}. Then we construct the block-encoding of $\frac{H_i'}{\gamma_i'} \otimes \Ibb_d \otimes \cdots \otimes \Ibb_d$, followed by the appropriate SWAP gates to obtain the original $\frac{H_i}{\gamma_i'}$. Then we use Lemma \ref{lemma: sumencoding} to build the block-encoding of $\frac{H}{\sum_{i=1}^K \gamma_i'}$, followed by an application of Lemma \ref{lemma: amp_amp} to remove the factor in the denominator, so as to obtain the block-encoding of $H$. Then finally we use Lemma \ref{lemma: qsvt} to obtain the block-encoding of $\exp(- i Ht )$.

The sparsity $s(\rho_j')$  of $\rho_j'$ is of the order $\mathcal{O}\big( d^{|R_i|} \big)$ in the worst case, which can be significantly smaller than $s(\rho_j)$ above, especially when $|R_i| \ll M$. In this case, the value of $s_\rho$ is bounded to the upper level as $\sum_{j=1}^N  s(\frac{\rho_j'}{N}) + s(\rho')  $, which can be considerably smaller than before. \\

Finally, in order to guaranty that the left-hand side of Eqn.~\ref{eqn: 18} is less than $\epsilon$, we need to have $ \Big|  \sum_{j=1}^N \frac{\rho_j}{N} - \rho    \Big|_{\infty} \leq \frac{\epsilon}{s_\rho}$. So, by virtue of the central limit theorem, it holds that:
\begin{align}
 \mathbb{P}\Big( \Big|  \sum_{j=1}^N \frac{\rho_j}{N} - \rho    \Big|_{o} \leq \epsilon \Big) = \mathbb{P}\Big( \Big|  \sum_{j=1}^N \frac{\rho_j}{N} - \rho    \Big|_{\infty} \leq \frac{\epsilon}{s_\rho} \Big) \geq 1- \frac{ s_\rho^2 \sigma^2}{N \epsilon^2}. 
\end{align}
So in this case, we need to choose $N = \mathcal{O}\big( \frac{s_\rho^2}{\epsilon^2} \big)$ to guaranty that this probability is sufficiently high. Because the number of samples $N$ depends on $s_\rho$, which is the sparsity of $\sum_{j=1}^N \frac{\rho_j}{N} - \rho= \sum_{j=1}^N \frac{\rho_j}{N} - \frac{H_i}{\gamma_i}$, and thus for each $H_i$ the value of $s_\rho$ is different. To make it more convenient for subsequent analysis, by a slight abuse of notation, we choose $N = \frac{s_\rho^2}{\epsilon^2}$ for all $H_1,H_2,...,H_K$ for $s_\rho$ now is defined as the maximum $s_\rho$ among all values of $s_\rho$'s (each with different $H_i$), or that $s_\rho \equiv \text{maximum sparsity of  \ }\{  \sum_{j=1}^N \frac{\rho_j}{N} - \frac{H_i}{\gamma_i} \}_{i=1}^K $. In the special cases discussed above, the value of $s_\rho$ is replaced by $s_\rho'$ and thus $N = \mathcal{O}\big(  \frac{s_\rho'^2}{\epsilon}\big)$.

As stated in the second step of Algo.~\ref{algo: timeindependentsimulation}, for any $j \in [1,2,...,N]$, the exact block-encoding of $\rho_j$ can be obtained. Then we can use Lemma \ref{lemma: sumencoding} to obtain the exact block-encoding of $\frac{1}{N} \sum_{j=1}^N \rho_j$ which is an $\epsilon$-approximation (in operator norm) to $\rho$ with high probability. We note that $\rho = \frac{H_j}{\gamma_i}$ so from this step we can proceed similarly to Algorithm \ref{algo: timeindependentsimulation} (from Step 3 onward).

\subsection{A third approach to prepare the block-encoding of $\varpropto H$  }
\label{sec: thirdapproach}
In the previous sections, we have described two approaches to block-encode $\frac{H_i}{\gamma_i}$. The idea underlying the second approach was to separately block-encode the projector
$$ \ket{\lambda_{1}(k_1)} \bra{ \lambda_{1}(k_1)}\otimes \ket{ \lambda_{2}(k_2)}  \bra{ \lambda_{2}(k_2)} \cdots \otimes \ket{\lambda_{M} (k_M)} \bra{ \lambda_{M} (k_M)},$$
and then using a linear combination of unitaries (with appropriate combination factors) to form the block-encoding of $\frac{H_i}{\gamma_i}$. At the same time, the idea behind the second approach was to sample a set of pure states with a classical probability distribution. Then we block-encode each $\rho_j$ with Lemma~\ref{lemma: improveddme} before we block-encode $\rho$ via Lemma~\ref{lemma: sumencoding}. 

In this section, we introduce another approach to achieve the given task, based on the recent result of~\cite{harrow2025randomized}. However, this approach requires that each $H_i$ be nonnegative. The general idea is based on the following observation. Consider the following state: 
\begin{align}
    \ket{\Phi}_i =  \frac{1}{\sqrt{\gamma_i}}  \sum_{\substack{k_1,k_2,...,k_M = 1 }}^d \sqrt{ \lambda_{i_1}(k_1) \lambda_{i_2}(k_2) \cdots \lambda_{i_M} (k_M)}  \underbrace{ \ket{k_1 k_2 .... k_M}}_{\rm extra \ register \ }\Big( \ket{\lambda_{i_1}(k_1)}\otimes \ket{ \lambda_{i_2}(k_2)} \cdots \otimes \ket{\lambda_{i_M} (k_M)} \Big),
    \label{eqn: 12}
\end{align}
where $\gamma_i  = \sum_{\substack{k_1,k_2,...,k_M = 1 }}^d |\lambda_{i_1}(k_1) \lambda_{i_2}(k_2) \cdots \lambda_{i_M} (k_M) |$ is the normalization factor. It can be seen that if we trace out the extra register in the above state, the resulting density state is:
\begin{align}
\begin{split}
     \rho = \frac{1}{\gamma_i}  \sum_{\substack{k_1,k_2,...,k_M = 1 }}^d  \lambda_{i_1}(k_1) \lambda_{i_2}(k_2) \cdots \lambda_{i_M} (k_M) \Big( \ket{\lambda_{i_1}(k_1)}\otimes \ket{ \lambda_{i_2}(k_2)} \cdots \otimes \ket{\lambda_{i_M} (k_M)} \Big) \\ \times \Big(\bra{\lambda_{i_1}(k_1)}\otimes \bra{ \lambda_{i_2}(k_2)} \cdots \otimes \bra{\lambda_{i_M} (k_M)} \Big), 
\end{split}
\end{align}
which is exactly $\frac{H_i}{\gamma_i}$. Therefore, if we can construct a unitary that prepares the state $\ket{\Phi}$, then we can combine it with Lemma \ref{lemma: improveddme} to prepare the block-encoding of the desired operator. In principle, since we already have the classical information of the eigenvectors $\{ \ket{\lambda_{i_1}(k_1)},  \ket{ \lambda_{i_2}(k_2)} \cdots , \ket{\lambda_{i_M} (k_M)}  \} $, we can use Lemma \ref{lemma: statepreparation} to prepare the state $\ket{\Phi}$. However, we note the following subtleties. In the previous approaches, we take the tensor product of the unitaries $U_{k_1} \otimes U_{k_2} \otimes \cdots \otimes U_{k_M}$ to obtain the unitary that prepares the state $ \ket{\lambda_{i_1}(k_1)} \otimes  \ket{ \lambda_{i_2}(k_2)} \cdots \otimes \ket{\lambda_{i_M} (k_M)} $.  We then need to use each unitary $ U_{k_1} \otimes U_{k_2} \otimes \cdots \otimes U_{k_M}$ separately combined with Lemma \ref{lemma: improveddme} (plus Lemma \ref{lemma: sumencoding}) to obtain the desired block-encodings. Here, in order to prepare $\ket{\Phi}$, we need to have another unitary which prepares the superposition as indicated in Eqn.~\ref{eqn: 12}. This means that we need to know all the entries in $\ket{\Phi}$ so that Lemma \ref{lemma: statepreparation} is applicable. This boils down to classical knowledge of all the (sub)vectors $\{ \ket{\lambda_{i_1}(k_1)}\otimes \ket{ \lambda_{i_2}(k_2)} \cdots \otimes \ket{\lambda_{i_M} (k_M)}  \}_{k_1,k_2,...,k_M= 1}^d $ and $i=1,2,...,K$. To achieve this, we need to classically perform the tensor product $\ket{\lambda_{i_1}(k_1)}\otimes \ket{ \lambda_{i_2}(k_2)} \cdots \otimes \ket{\lambda_{i_M} (k_M)}  $ for all values of $k's,i$. However, this naive approach is definitely costly, because each vector composing has dimension $d$, so the total multiplication time to obtain the tensor product of this type is $\mathcal{O}(d^M) = \mathcal{O}(2^n)$. One scenario that can improve this multiplication complexity is when all states $ \ket{\lambda_{i_1}(k_1)},  \ket{ \lambda_{i_2}(k_2)} \cdots , \ket{\lambda_{i_M} (k_M)} $ have a bounded sparsity (ideally much smaller than $d$), denoted by $s$. Then the complexity of performing this tensor product is $\mathcal{O}(s^M)$ (subsequently we will discuss that this complexity can be even much smaller than this value in special cases in discussed in Sec.~\ref{sec: simplification}), which can be significantly less than $\mathcal{O}(2^n)$. However, in reality, the sparsity depends on the structure of each $H_i$, and thus there is no general guarantee that the sparsity is small. 

To this end, we point out that recently, Ref.~\cite{harrow2025randomized} introduced a method to approximately decompose a pure state (or more precisely, the density matrix corresponds to a pure state) into an ensemble of sparse states. Their method, named randomized truncation to a quantum state, is summarized in the following lemma:
\begin{lemma}[Randomized approximation to a quantum state \cite{harrow2025randomized}]
    \label{lemma: randomapproximation}
Let $v\in \mathbb{C}^d$ be a unit vector, e.g., $||v||_2 = 1$ and $s \leq d$ be some integer. Given the knowledge of entries of $v$, there exists a classical algorithm running in time $\rm poly (d)$ to find a classical description of $\{ p_i , w_i\}_{i=1}^L$ with $\{p_i\}_{i=1}^L$ being some probability distribution (which can be efficiently sampled from) and for all $i=1,2,...,L$, $w_i \in \mathbb{C}^d$ is some $s$-sparse unit vector, such that the trace distance  
$$ \Tr \Big|  v v^\dagger - \sum_{j=1}^L p_j w_j w_j^\dagger \Big|  $$
is optimally minimized. 
\end{lemma}
The above results essentially provide a classical procedure that ``breaks'' the density state $v v^\dagger $ (with $v$ being a known state) into an ensemble of pure and sparse states $\{ w_i\}_{i=1}^L$ with appropriate distribution. A particular feature of the classical algorithm above is that the value of $s$ can be chosen, and for different values of $s$, the trace distance is accordingly optimized. In particular, the complexity of such a procedure does not depend on the approximation error, as quantified by the trace distance. Therefore, we can safely treat the approximation error to be $\mathcal{O}(1)$ and it will not (asymptotically) affect the overall complexity of the algorithm. 

To apply the result above to our context, i.e., to prepare $|\Phi_i\rangle$'s as in Eqn.~\ref{eqn: 12}, we consider the state $\ket{\lambda_{i_1}(k_1)}$. Since the classical knowledge of this state is known (via the classical diagonalization of $H_i$), the Lemma above allows us to find those $s$-sparse states $ w_{1}(i_1,k_1), w_{2}(i_1,k_1), w_{3}(i_1,k_1), ..., w_{L}(i_1,k_1) \}$ which approximate $\ket{\lambda_{i_1}(k_1)}\bra{\lambda_{i_1}(k_1)}$.  The same lemma can be employed to find those states that approximate $ \ket{\lambda_{i_2}(k_2)}, ..., \ket{\lambda_{i_M}(k_M}$. We then observe the following property. Consider the state:
\begin{align}
    \sum_{j=1}^L \sqrt{p_j} \ket{j} w_j(i_1,k_1).
\end{align}
If we trace out the first register (the one that contains $\ket{j}$), then we obtain the density state $\sum_{j=1}^L p_j w_j w_j^\dagger$ which is close to $\ket{\lambda_i(k_1)}\bra{\lambda_i(k_1)}$. Now we consider the following state (with $\ket{j_1 j_2} \equiv \ket{j_1}\ket{j_2}$):
\begin{align}
    \sum_{j_1, j_2 =1}^L \sqrt{p_{j_1} p_{j_2}}  \ket{j_1 j_2}  w_{j_1}(i_1,k_1) \otimes  w_{j_2} (i_2,k_2).
\end{align}
If we trace out the register that holds the state $\ket{j_1 j_2}$, then we obtain the following state: 
\begin{align}
    \Big(  \sum_{j_1=1}^L p_{j_1} w_{j_1}(i_1,k_1)  w_{j_1}(i_1,k_1)^\dagger  \Big) \otimes \Big( \sum_{j_2=1}^L p_{j_2} w_{j_2}(i_2,k_2) w_{j_2}(i_2,k_2)^\dagger \Big),
\end{align}
which is close to the state $\ket{\lambda_{i_1} (k_1)} \bra{\lambda_{i_1}(k_1)} \otimes \ket{\lambda_{i_2}(k_2)} \bra{\lambda_{i_2)}(k_2}$. Now we consider the following state:
\begin{align}
\begin{split}
 \frac{1}{\sqrt{\gamma_i}} \sum_{\substack{k_1,k_2,...,k_M = 1 }}^d \sqrt{ \lambda_{i_1}(k_1) \lambda_{i_2}(k_2) \cdots \lambda_{i_M} (k_M)} \underbrace{  \ket{k_1 k_2 .... k_M}}_{\rm extra \ register \ 1 } \sum_{j_1,j_2,...,j_M=1}^L \Big( \sqrt{ p_{j_1} p_{j_2} ... p_{j_M} } \underbrace{ \ket{j_1 j_2 ... j_M } }_{\rm extra \ register \ 2}\otimes w_{j_1} (i_1,k_1) \\ \otimes w_{j_2}(i_2,k_2) \otimes \cdots \otimes \omega_{j_M} (i_M, k_M)\Big).    
\end{split}
\label{eqn: 17}
\end{align}
Generalizing from the earlier examples, it can be seen that once we trace out the extra register 1 and 2 from the above state, we obtain the following density state:
\begin{align}
\begin{split}
     \frac{1}{\gamma_i} \sum_{\substack{k_1,k_2,...,k_M = 1 }}^d i^{ \{0/1\}}   \lambda_{i_1}(k_1) \lambda_{i_2}(k_2) \cdots \lambda_{i_M} (k_M)\Big(  \sum_{j_1=1}^L p_{j_1} w_{j_1}(i_1,k_1)  w_{j_1}(i_1,k_1)^\dagger  \Big) \otimes  \Big( \sum_{j_2=1}^L p_{j_2} w_{j_2}(i_2,k_2) w_{j_2}(i_2,k_2)^\dagger \Big) \\ \otimes  \Big( \sum_{j_M=1}^L p_{j_M} w_{j_M}(i_M,k_M) w_{j_M}(i_M,k_M)^\dagger \Big),
     \label{26}
\end{split}
\end{align}
which is an approximation to $\rho' = \frac{H_i}{\gamma_i}$. Since all the entries of the states $\{   w_{j_1} (i_1,k_1) , w_{j_2}(i_2,k_2) , \cdots , \omega_{j_M} (i_M, k_M) \}_{j_1,j_2,..,j_M =1}^L $ are known, then the state $\ket{\Phi'}$ above can be prepared using Lemma \ref{lemma: statepreparation} with complexity $ \mathcal{O}\Big( \log \big ( L^M \sum_{i=1}^K r(H_i) s^M\big) \Big) $. The density state $\rho'$ above can also be block-encoded via Lemma \ref{lemma: improveddme}. Once we have the block-encoding of $\frac{H_i}{\gamma_i}$, we can execute similarly to the Algorithm \ref{algo: timeindependentsimulation} (from Step 3 onward). Then, we can use Lemma \ref{lemma: qsvt} with the Jacobi-Anger expansion to transform the block-encoding of $H$ to the block-encoding of $\exp\big( - i H t \big)$ as the final step. \\

\noindent
\textbf{Simplification in special cases (following Section \ref{sec: simplification}). } Again, in the special cases as discussed in Section \ref{sec: simplification}, if there are many trivial terms within each $H_i$, then we can define a new Hamiltonian $H_i'$ and repeat the above algorithm to block-encode $\varpropto H_i'$, followed by block-encoding $\varpropto H_i' \otimes \Ibb_d \otimes \cdots \otimes \Ibb_d $ and multiply it with the (block-encoded) SWAP gate to recover the original $\varpropto H_i$. The remaining steps are the same as we use Lemma \ref{lemma: sumencoding} to obtain the block-encoding of $\varpropto H$ and finally $\exp(-iHt)$. 

\subsection{Generalization to time-dependent but commuting Hamiltonians}
\label{sec: generalization}

In this section, we discuss how our hybrid algorithm above can be extended to time-dependent setting. Generally, time-dependent Hamiltonian simulation is significantly more challenging compared to the time-independent setting. The reason is that in the time-dependent regime, the evolution operator is no longer as simple as $\exp\big(-i H t \big)$, but rather it is $\tau\exp\big(-i \int_0^t H(s)ds \big)$ where $\tau$ is the time-ordered operator. Therefore, in the general case, the evolution operator does not admit an analytical form, which incurs nontrivial difficulty for the simulation. However, a particular scenario that dramatically simplifies things is when the Hamiltonian $H(t)$ commutes at different times, e.g., for $t_1 \neq t_2$, it holds that $H(t_1) H(t_2) = H(t_2) H(t_1)$. In this case, the evolution operator admits an analytical expression, e.g., $\exp\big( -i \int_0^t H(s) ds\big)$, which removes the time-ordered operator. Therefore, as long as $H(t)$ is efficiently integrable, the term in the exponent can be analytically written down. 

We recall that the Hamiltonian model that we have worked with has the following form $H = \sum_{i=1}^K H_i = \sum_{i=1}^K H_{i_1} \otimes H_{i_2} \otimes \cdots \otimes H_{i_M}$. The time-dependent Hamiltonian that we would be interested in is modified from this model as follows:
\begin{align}
    H(t) = \sum_{i=1}^K \alpha_i(t) H_i = \sum_{i=1}^K  \alpha_i(t) H_{i_1} \otimes H_{i_2} \otimes \cdots \otimes H_{i_M},
\end{align}
where $\{\alpha_i(t)\}_{i=1}^K$ are efficiently integrable functions of time $t$. To have the Hamiltonian above commutes at different times, the condition we need is that an arbitrary pair among $\{H_i\}_{i=1}^K$ commutes. The evolution operator of this Hamiltonian then is:
\begin{align}
    \exp\big( -i \int_0^t H(s) ds \big) = \exp\big( - \sum_{i=1}^K (\int_0^t\alpha_i(s)ds) H_i  \big). 
\end{align}
Our hybrid algorithm to simulate $H(t)$ in this case proceeds almost similarly to the one in the previous section, as we still need to perform Step 1 through Step 5. The only difference is that we do not use Lemma \ref{lemma: qsvt} to directly obtain the simulation operator as in Step 6. Rather, we proceed as follows. 
\begin{method}
\label{algo: timedependentsimulation}
    The algorithm for simulating time-dependent $H= \sum_{i=1}^K  \alpha_i(t) H_{i_1} \otimes H_{i_2} \otimes \cdots \otimes H_{i_M}$ proceeds as follows:
\begin{enumerate}
    \item Repeat Step 1-4 as in Algo.~\ref{algo: timeindependentsimulation}, we obtain the $\epsilon$-approximated block-encoding of $\{ \frac{H_i}{\gamma_i}\}_{i=1}^K$. 
    \item Consider the following single-qubit rotation gate: 
    \begin{align}
        \exp(-i\sigma_Z \theta) = \begin{pmatrix}
            \cos \frac{\theta}{2}  & i \sin \frac{\theta}{2} \\
            i \sin \frac{\theta}{2} & -\cos \frac{\theta }{2}
        \end{pmatrix}.
    \end{align}
 Choosing $\frac{1}{2}\theta = \arccos(t)$, the above gate is as follows:
    \begin{align}
        \begin{pmatrix}
            t & i \sqrt{1-t^2} \\
            i \sqrt{1-t^2} & -t
        \end{pmatrix}.
    \end{align}
    \item We define $ \int_0^t \alpha_i(s)ds =  \beta_i(t)$. As each $\alpha_i(t)$ was assumed to be integrable efficiently, $\beta_i(t)$ can be obtained efficiently. Now we perform either the quantum signal processing technique (as in Ref.~\cite{low2017optimal}) or Lemma \ref{lemma: qsvt} with the polynomial $P \equiv \beta_i$, to transform the above gate (which is a unitary operator and block-encodes itself):
    \begin{align}
        \begin{pmatrix}
            t & i \sqrt{1-t^2} \\
            i \sqrt{1-t^2} & -t
        \end{pmatrix} \longrightarrow \begin{pmatrix}
            \beta_i(t)  & \cdot \\
            \cdot & \cdot 
        \end{pmatrix},
    \end{align}
    where the $(\cdot )$ refers to the irrelevant part, as we only pay attention to the top-left entry (the block-encoded one) $\beta_i(t)$. 
    \item We use Lemma \ref{lemma: tensorproduct} with the block-encoding of the above operator and of the operator $\frac{H_i}{\gamma_i}$ obtained from the first step, to obtain the block-encoding of:
    \begin{align}
        \begin{pmatrix}
            \beta_i(t)  & \cdot \\
            \cdot & \cdot 
        \end{pmatrix} \otimes \frac{H_i}{\gamma_i} = \begin{pmatrix}
            \beta_i(t) \frac{H_i}{\gamma_i} & \cdot \\
            \cdot & \cdot 
        \end{pmatrix},
    \end{align}
    which is exactly the block-encoding of $ \beta_i(t) \frac{H_i}{\gamma_i}$. 
    \item Repeat Step 5 of the Algo.~\ref{algo: timeindependentsimulation} to construct the $\epsilon$-approximated block-encoding of 
    \begin{align}
        \sum_{i=1}^K \frac{\gamma_i}{ \sum_{i=1}^K\gamma_i} \frac{\beta_i(t) H_i}{\gamma_i} = \frac{1}{\sum_{i=1}^K\gamma_i} \sum_{i=1}^K \int_0^t \alpha_i(s)ds H_i.
    \end{align}
    \item Next, we use Lemma \ref{lemma: amp_amp} to remove the factor $\sum_{i=1}^K \gamma_i$ out of the above block-encoded operator. 
    \item The final step is to transform the above block-encoded operator into $ \exp\big( - \sum_{i=1}^K (\int_0^t\alpha_i(s)ds) H_i  \big)  $, which can be done by employing Lemma \ref{lemma: qsvt} with the Jacobi-Anger expansion as in Step 6 of Algo.~\ref{algo: timeindependentsimulation}, i.e., transform the block-encoded operator:
    \begin{align}
       \sum_{i=1}^K \int_0^t \alpha_i(s)ds H_i \longrightarrow \exp\Big( -i \big( \sum_{i=1}^K \int_0^t \alpha_i(s)ds H_i \big) t'\Big).
    \end{align}
    By choosing $t'$, we obtain the block-encoding of $ \exp\Big( - i\sum_{i=1}^K \big( \int_0^t\alpha_i(s)ds \big) H_i  \Big)$, which is exactly the desired evolution operator. 
\end{enumerate}
\end{method}

As a brief comment, our algorithm above is limited to the case where terms $\{ H_i\}_{i=1}^K$ commute, for which we can take advantage of the analytical expression of the evolution operator. For the noncommuting case, one could naively prepare $e^{-i H(t) \Delta t}$ (for sufficiently small $\Delta t$) with our method in different $t$ to approximate the time-ordered evolution $\prod_i e^{-i H(t_i) \Delta t}$ (similar to the Lie-Trotter-Suzuki formula). Despite being straightforward, we suspect this approach may not be very efficient, and there will be significant scaling in the inverse of the error tolerance due to error accumulation from Trotter steps.

\smallskip
\noindent
\textbf{Simplification in special cases.} The above algorithm can apparently be simplified in an exact manner as we discussed in Section~\ref{sec: simplification}. When there are many $H_{i_j}$'s are identity matrices,  we just need to define a new matrix (for each $i$) $H_i'$ and then repeat the above algorithm to block-encode $\varpropto H_i'$ and then eventually $\varpropto H$. 

\section{Complexity analysis}
\label{sec: complexityanalylsis}
In this section, we first analyze in detail the circuit complexity of Algorithm \ref{algo: timeindependentsimulation}, followed by the analysis for the alternative approach in Sec.~\ref{sec: alternativeapproach}. Then we analyze the complexity of Algorithm \ref{algo: timedependentsimulation}.  

\subsection{Analysis of time-independent simulation algorithm Algo.~\ref{algo: timeindependentsimulation}}
We will follow the hybrid algorithm, including the classical component (Sec.~\ref{sec: classical}) and quantum component, e.g., Algorithm \ref{algo: timeindependentsimulation} (Sec.~\ref{sec: quantum}) step by step. 
\begin{itemize}
\item[$\blacktriangleright$]First, regarding the classical component, we need to use a classical algorithm to diagonalize the matrices $\{ H_{i_j} \}_{j=1}^M $ for all $i$. As the dimension of each $H_{i_j}$ is $d \times d$, the classical complexity is then $\mathcal{O}(d^3)$, which can be improved to $\mathcal{O}(sd)$ if all $H_{i_j}$ is sparse, with $s$ being the sparsity. 
    \item[$\blacktriangleright$] In Step 1, for a given $i \in [1,2,...,K]$, we need to use Lemma~\ref{lemma: statepreparation} to obtain the unitaries $U_{k_1},U_{k_2},...,U_{k_M}$ that prepare the states $\ket{\lambda_{i_1}(k_1)},  \ket{ \lambda_{i_2}(k_2)} \cdots , \ket{\lambda_{i_M} (k_M)} $. Each state $\ket{\lambda_{i_j}(k_j)}$ has dimension $d$, and the sparsity is at most $d$, so the circuit complexity (in terms of gate) of each $U_{k_j}$ is $\mathcal{O}\big( \log d \big)$ plus $\mathcal{O}(d)$ ancilla qubits, which means that the circuit complexity for $ \bigotimes_{j=1}^M U_{k_j}$ is $\mathcal{O}\big( M\log d \big)$ plus extra $\mathcal{O}( Md)$ ancilla qubits. However, we note that the depth is still $\mathcal{O}\big( \log d \big)$ because all the unitaries $\{U_{k_j}\}_{j=1}^M$ are placed in parallel. 
    
    \item[$\blacktriangleright$] In Step 2 and 3, we use Lemma \ref{lemma: improveddme} to first prepare the exact block-encoding of 
$$  \{ \ket{\lambda_{i_1}(k_1)}\otimes \ket{ \lambda_{i_2}(k_2)} \cdots \otimes \ket{\lambda_{i_M} (k_M)} \ \bra{\lambda_{i_1}(k_1)}\otimes \bra{ \lambda_{i_2}(k_2)} \cdots \otimes \bra{\lambda_{i_M} (k_M)}\} _{\substack{k_1,k_2,...,k_M = 1 }}^d. $$
Lemma~\ref{lemma: improveddme} uses $\mathcal{O}(1)$ unitary $\bigotimes_{j=1}^M U_{k_j}$ (each term in the tensor product has circuit complexity $\mathcal{O}(\log d)$ plus $\mathcal{O}(d)$ ancilla qubits and another $\mathcal{O}\big( M \log d \big)$ 2-qubit gates (as the dimension of the operator above is $d^M$), so the total circuit complexity for the block-encoding above is $\mathcal{O}(M\log d )$ and extra $\mathcal{O}(Md)$ ancilla qubits. Then, we use Lemma~\ref{lemma: sumencoding} to obtain the linear combination (with combination factors $\{  \frac{ \lambda_{i_1}(k_1) \lambda_{i_2}(k_2) \cdots \lambda_{i_M} (k_M)}{\gamma_i} \}_{\substack{k_1,k_2,...,k_M = 1 }}^d $) as in Eq.~(\ref{eqn: 11}), which is also $\frac{H_i}{\gamma_i}$. The total of terms in such a summation is $r(H_{i_1}) r(H_{i_2}) \cdots r(H_{i_M}) = r(H_i)$, where $r(.)$ denotes the rank of the corresponding matrix. So, the complexity in preparing the block-encoding of the operator $\rho$ is $\mathcal{O}\Big( r(H_{i_1}) r(H_{i_2}) \cdots r(H_{i_M}) (M \log d)    \Big)$.  
We note that as $ r(H_{i_1}) r(H_{i_2}) \cdots r(H_{i_M}) = r(H_i)$, the  complexity is then $\mathcal{O}\Big( r(H_i) (M \log d)    \Big). $

    \item[$\blacktriangleright$] In Step 4, we need to repeat the above 3 steps $K$ times obtain the $\epsilon$-approximated block-encoding of $\{ \frac{H_i}{\gamma_i}\}_{i=1}^K$. The circuit complexity for each of them is $\mathcal{O}\Big( r(H_i) (M \log d)    \Big) $.
    
    \item[$\blacktriangleright$] In Step 5, we need to use Lemma~\ref{lemma: sumencoding} to form the linear combination that leads to the $\epsilon$-approximated block-encoding of $\frac{H}{\sum_i \gamma_i}$. The linear combination uses each block-encoding of $\{\frac{H_i}{\gamma_i}\}_{i=1}^K $, so the total circuit complexity is 
    $$\mathcal{O}\Big( \sum_{i=1}^K  r(H_i) (M \log d)    \Big). $$
    
    \item[$\blacktriangleright$] In Step 6, we need to use Lemma~\ref{lemma: amp_amp} to remove the factor $\sum_{i=1}^K \gamma_i$ from the above block-encoded operator. Lemma~\ref{lemma: amp_amp} uses the block-encoding above a total of $\mathcal{O}\left( |\sum_{i=1}^K \gamma_i | \right)$ times, resulting in a total complexity 
     $$\mathcal{O}\Big(  \sum_{i=1}^K \gamma_i \sum_{i=1}^K  r(H_i) (M \log d)    \Big). $$
    
    \item[$\blacktriangleright$] In Step 7, we need to use Lemma~\ref{lemma: qsvt} with polynomial $P$ being the Jacobi-Anger expansion as used in Ref.~\cite{low2017optimal, low2019hamiltonian}. To approximate $\exp(-i x t) $ on the interval $x\in [-1,1] $ with a precision $\delta$, the degree of this polynomial is $\mathcal{O}\left( t \log \frac{1}{\delta}    \right)$. So, the total circuit complexity is 
     $$\mathcal{O}\Big(  \sum_{i=1}^K \gamma_i  \sum_{i=1}^K  r(H_i) (M \log d)     t \log \frac{1}{\delta}\Big). $$
    
\end{itemize}
Therefore, we arrive at the final circuit complexity (again $\mathcal{\tilde{O}}$ hides the polylog terms), including both the classical and quantum part:
$$      \mathcal{O}\Big(  \sum_{i=1}^K \gamma_i  \sum_{i=1}^K  r(H_i) (M \log d)     t \log \frac{1}{\delta}\Big).         $$
and the total number of extra ancilla qubits is $\mathcal{O}(Md)$. 

As we discussed in Section \ref{sec: simplification}, the quantum complexity can be reduced further when many of the matrices $H_{i_1}, H_{i_2}, ..., H_{i_M}$ are easily block-encoded operator, e.g., identity operators. In this case, it can be seen that we only need to diagonalize those remaining operators (which are not easily block-encoded) to obtain the spectrum before we block-encode them via a combination of Lemma \ref{lemma: statepreparation} and \ref{lemma: improveddme}. As such, for a given $i$, we are effectively dealing with the Hamiltonian $H_i' = \otimes_{H_{i_j} \in R_i} H_{i_j}$. Therefore, for each $i$, the value of $M$ in the complexity above is reduced to $|R_i|$, the rank $r(H_i)$ in the above complexity would be reduced to
\begin{align}
    r(H_i') = \prod_{H_{i_j} \in R_i} r (H_{i_j}).
\end{align}
Additionally, the value of $\gamma_i$ is reduced to
$$\gamma_i' = \sum_{\substack{k_1,k_2,...,k_{|R_i|} = 1 }}^d |\lambda_{i_1}(k_1) \lambda_{i_2}(k_2) \cdots \lambda_{i_{|R_i| }} (k_{|R_i|}) |.$$

Further, we note that for each $i$, at the Step 4 of Algo \ref{algo: timeindependentsimulation},  once we obtain the block-encoding of $\frac{H_i'}{\gamma_i'}$ (with complexity $\mathcal{O}\big( r(H_i') (|R_i| \log d)  \big)$), we need to construct the block-encoding of $\frac{H_i'}{\gamma_i'} \otimes \Ibb_d \otimes \cdots \otimes \Ibb_d$. Then via Lemma \ref{lemma: product}, we multiply such the block-encoded operator with the (block-encoded) SWAP operators to recover the original $\frac{H_i}{\gamma_i'}$. For each $i$, the maximum number of SWAP operations required is $|R_i|$ (as there are at most $|R_i|$ nontrivial terms need to be swapped), so the complexity is of order $\mathcal{O}( |R_i|)$. Adding this complexity to the complexity at Step 4 above, we have that the complexity for obtaining the block-encoding of $\frac{H_i}{\gamma_i'}$ is asymptotically the same, $\mathcal{O}\big( r(H_i) |R_i| \log d  \big)$. The complexity for the remaining steps are thus asymptotically the same, which is
$$ \mathcal{O}\Big(  \sum_{i=1}^K \gamma_i'  \big(\sum_{i=1}^K  \prod_{H_{i_j} \in R_i} r (H_{i_j}) \big)|R_i| \log (d)   t \log \frac{1}{\delta}\Big).  $$ 

Now we analyze more in detail the value of $\gamma_i'$ and $\prod_{H_{i_j} \in R_i} r (H_{i_j}) $. At the beginning, we have assumed that $||H_i||_o \leq 1/2$ for each $i$. So, for each $\gamma_i'$, which is defined as the sum of  absolute magnitude of the eigenvalues of $H_i'$, would have the magnitude of order  $\mathcal{O}(  r(H_i')  ) = \mathcal{O}\Big( \prod_{H_{i_j} \in R_i} r (H_{i_j})\Big)$. Each $H_{i_j}$ has dimension $d$, which means that its rank $r(H_{i_j})$ is at most $d$, so this product has magnitude $\mathcal{O}\big( d^{|R_i|} \big)$. So the complexity above is further bounded by:

$$ \mathcal{O}\Big(   \sum_{i=1}^K  d^{|R_i|} \big(\sum_{i=1}^K d^{|R_i|} \big) |R_i| \log (d)   t \log \frac{1}{\delta}  \Big).$$

Defining $|R| \equiv \max\{  |R_i| \}$, then we have that for all $i$,  $d^{|R_i|} \leq d^{|R|}$. So the complexity above can be reduced to 
$$  \mathcal{O}\Big(  K^2 d^{2|R|}   |R| \log (d)     t \log \frac{1}{\delta}\Big), $$
where we have used the assumption that $||H_i||_o \leq 1/2$ for each $i$.

\subsection{Analysis of the alternative approach in Sec.~\ref{sec: alternativeapproach}}
As described in Section~\ref{sec: alternativeapproach}, the approach of this section differs only from the algorithm in Section~\ref{sec: quantum} in the second and third steps. Instead of preparing the block-encoding of $\rho = \frac{H_i}{\gamma_i}$, we prepare the block-encoding of $\sum_{j=1}^N \frac{\rho_j}{N}$, which is $\epsilon$-close to $\rho$ (in operator norm) with high probability. Each $\rho_j$ is block-encoded by first randomly selecting the state among $ \ket{\lambda_{i_1}(k_1)}\otimes \ket{ \lambda_{i_2}(k_2)} \cdots \otimes \ket{\lambda_{i_M} (k_M)}$ with probability $p_i(k_1,k_2,...,k_M) $. In addition to the classical time required for diagonalization, there is also a classical sampling step that produces $N$ samples $\rho_1,\rho_2,...,\rho_N$ and we need to repeat for $H_1,H_2,...,H_K$ so the total classical time is $\mathcal{O}(NK)$. 

As noted in the previous section, the circuit complexity for the encoding of each block $\rho_j$ (which is essentially some $\ket{\lambda_{i_1}(k_1)}\otimes \ket{ \lambda_{i_2}(k_2)} \cdots \otimes \ket{\lambda_{i_M} (k_M)} \ \bra{\lambda_{i_1}(k_1)}\otimes \bra{ \lambda_{i_2}(k_2)} \cdots \otimes \bra{\lambda_{i_M} (k_M)}$) is $\mathcal{O}\big(M \log d\big)$. So, the total circuit complexity for the block-encoding $\sum_{j=1}^N \frac{\rho_j}{N} $ is $\mathcal{O}\Big( N (M \log d) \Big)$. The remaining steps are the same as in Algo.~\ref{algo: timeindependentsimulation}, as we build the block-encoding of $H$, which involves building the linear combination of $\{ \frac{H_i}{\gamma_i} \}_{i=1}^K$, incurring a total circuit complexity 
$$ \mathcal{O}\Big( N K (M \log d) \Big).$$
Then we use the QSVT with Jacobi-Anger expansion polynomial of degree $\mathcal{O}(t \log \frac{1}{\epsilon})$ to obtain the block-encoding of $\exp(-i H t)$. So, the total complexity for the quantum part can be shown to be
$$ \mathcal{O}\Big(  NK \big(  \sum_{i=1}^K \gamma_i  \big)  (M \log d)   t \log \frac{1}{\epsilon}  \Big),  $$
and the total number of extra ancilla qubits is $\mathcal{O}(Md)$.  

However, there is a subtlety that we would like to point out. The block-encoding of $ \sum_{j=1}^N \frac{\rho_j}{N}$ is $\epsilon$-close (in the operator norm) to $\rho$. Therefore, the block-encoding of  $H$ is actually $\epsilon$-approximated. In the Lemma~\ref{lemma: qsvt}, we have that when performing the QSVT with a polynomial of degree $P$, we have that the final error of the block-encoded operator (in the operator norm) is $\deg(P) \sqrt{\epsilon}$, which is $ \sqrt{\epsilon}\, t \log \frac{1}{\epsilon}$. In order for the overall error of the block-encoding of $\exp(-i H t)$ to be $\delta$, we need to require that $\sqrt{\epsilon}\, t \log \frac{1}{\epsilon} = \delta$, which implies that $\epsilon = \mathcal{O}\big(  \frac{\delta^2}{t^2} \big)$. 

In Section~\ref{sec: alternativeapproach}, we have seen that the value of $N$ is $\mathcal{O}\big(  \frac{s_\rho^2}{\epsilon^2} \big)$ for $s_\rho \equiv \text{maximum sparsity of  \ }\{  \sum_{j=1}^N \frac{\rho_j}{N} - \frac{H_i}{\gamma_i} \} $. the total complexity is 
$$ \mathcal{O}\Big(  K \big(  \sum_{i=1}^K \gamma_i  \big)  (M \log d)  \frac{s_\rho^2}{\epsilon^2} t \log \frac{ t}{\delta}  \Big),$$
which is 
$$  \mathcal{O}\Big(  K \big(  \sum_{i=1}^K \gamma_i  \big)  (M \log d)  \frac{s_\rho^2}{\delta^2} t^3 \log \frac{1}{\delta}  \Big).     $$
As we discussed in  Section~\ref{sec: alternativeapproach}, the value of $s_\rho$ can be as large as $\mathcal{O}(d^M)$, which can be reduced to $\mathcal{O}\big( d^{|R_i|}  \big)$ in the special case where there are trivial matrices among $\{ H_{i_j} \}_{j=1}^M$ (for each $i$). In this case, as we have discussed in the previous section, we need to replace $\gamma_i \longrightarrow \gamma_i'$, $ M \longrightarrow |R|$. Therefore, the complexity above can be reduced to 
$$ \mathcal{O}\Big(  K \big(  \sum_{i=1}^K \gamma_i ' \big) |R| \log (d)  \frac{d^{2|R_i|}}{\delta^2} t^3 \log \frac{1}{\delta}  \Big) = \mathcal{O}\Big(  K^2 |R| \log (d) \frac{ d^{4|R|}}{\delta^2} t^3 \log \frac{1}{\delta}  \Big), $$
where we have used that $  \sum_{i=1}^K \gamma_i ' = \mathcal{O}\big( \sum_{i=1}^K  d^{|R_i|} \big) = \mathcal{O}\big( K d^{|R|}  \big)$.

\subsection{Analysis of the third approach in Sec.~\ref{sec: thirdapproach}  }
This approach replaces the second, third, fourth and fifth steps of Algorithm~\ref{algo: timeindependentsimulation}, which aims to produce the block-encoding of $H$. As described in Sec.~\ref{sec: thirdapproach}, we first need to use Lemma~\ref{lemma: randomapproximation} to find the set of $s$-sparse states that achieve the desired approximation. For a state of dimension $d$, the complexity of Lemma~\ref{lemma: randomapproximation} is $\mathcal{O}\big( \rm poly (d) \big)$. In our case, we apply Lemma~\ref{lemma: randomapproximation} to the states of the form $\ket{\lambda_{i_1}(k_1)}\otimes \ket{ \lambda_{i_2}(k_2)} \cdots \otimes \ket{\lambda_{i_M} (k_M)}  $, so the classical complexity accumulated  is $\mathcal{O}\big( M \rm poly (d)\big)$. There are a total of $r(H_{i_1}) r(H_{i_2}) \cdots r(H_{i_M}) = r(H_i) $ terms in the summation $\sum_{k_1,k_2,...,k_M=1}^d$ and there are $K$ different values of $i$, so the total classical complexity of using Lemma~\ref{lemma: randomapproximation} is $\mathcal{O}\big( \sum_{i=1}^K r(H_i) M \rm poly (d) \big) $. However, there is one step where we need to perform the multiplication 
$$ w_{j_1} (i_1,k_1)  \otimes w_{j_2}(i_2,k_2) \otimes \cdots \otimes \omega_{j_M} (i_M, k_M).$$
Each state in the above has sparsity $s$, so the multiplication time is $\mathcal{O}(s^M)$, resulting in the total classical complexity 
$$\mathcal{O}\Big( s^M \sum_{i=1}^K r(H_i) M  \rm poly (d) \Big). $$

Regarding the complexity for the quantum part, we recall that we need to use Lemma~\ref{lemma: statepreparation} to prepare the state $\ket{\Phi}_i$, see, e.g., Eqn.~\ref{eqn: 17}. This state has dimension $ L^M r(H_{i_1}) r(H_{i_2}) \cdots r(H_{i_M})d^M  = L^M r(H_i) d^M$. However, each of the states $w$'s is actually $s$-sparse, so the total number of nonzero entries in the state $\ket{\Phi}_i$ is actually $ L^M r(H_i) s^M $. According to Lemma~\ref{lemma: statepreparation}, the quantum circuit depth complexity is 
$$ \mathcal{O}\Big( \log \big ( L^M  r(H_i) s^M\big) \Big), $$
and the required number of qubits is $ \mathcal{O}\left(  L^M r(H_i)s^M  \right)$ qubits, including ancillas.  The next step is to use Lemma~\ref{lemma: improveddme} to block-encoding $\rho \equiv H_i/\gamma_i$, which incurs the (asymptotically) same complexity as above. Subsequently, we use Lemma~\ref{lemma: sumencoding} to block-encode the $H/ \sum_{i=1}^K \gamma_i$, which uses the block-encoding of each $  H_i/\gamma_i$ one time. So, the total quantum circuit complexity is
$$ \mathcal{O}\Big( \sum_{i=1}^K \log \big ( L^M  r(H_i) s^M\big)  \Big)  = \mathcal{O}\Big( \log \big ( L^M \sum_{i=1}^K r(H_i) s^M\big)   \Big), $$
where the number of qubits usage is $  \mathcal{O}\left(  L^M \sum_{i=1}^K r(H_i)s^M  \right)$. The final steps are to use Lemma~\ref{lemma: amp_amp} to remove the factor in the denominator, which incurs a total depth complexity $\mathcal{O}\Big(  \sum_{i=1}^K \gamma_i\log \big ( L^M \sum_{i=1}^K r(H_i)s^M\big)  \Big)$ (the ancilla qubits remain asymptotically the same). Lastly, we use Lemma~\ref{lemma: qsvt} with the polynomial of degree $\mathcal{O}( t \log \frac{1}{\epsilon})$ so the final (depth) complexity is: 
$$ \mathcal{O}\Big( t  \sum_{i=1}^K \gamma_i \log \big ( L^M \sum_{i=1}^K r(H_i) s^M\big) \log \frac{1}{\epsilon}  \Big),$$
while using totally $  \mathcal{O}\left(  L^M s^M \sum_{i=1}^K r(H_i)   \right)$ ancilla qubits. 

We remark that, in the above, we have not accounted for the error approximation using Lemma \ref{lemma: randomapproximation}, which should propagate to the final error in simulating $H$. To remind, we have approximated, for example, the density operator
\begin{align}
    \ket{\lambda_{i_1} (k_1)}  \bra{\lambda_{i_1} (k_1)} \approx \sum_{j=1}^L p_j w_j(i_1,k_1) w_j(i_1,k_1)^\dagger.
\end{align}
For convenience, we define the following: 
\begin{align}
    \Delta = \max  \{ \Tr \Big|  \ket{\lambda_{i_1} (k_1)}  \bra{\lambda_{i_1} (k_1)} - \sum_{i=1}^L p_j w_j(i_1,k_1) w_j(i_1,k_1)^\dagger \Big| , ... ,  \Tr \Big| \ket{\lambda_{i_M} (k_M)}  \bra{\lambda_{i_M} (k_M)} - \sum_{j=1}^L p_j w_j(i_M,k_M) w_j(i_M,k_M)^\dagger \Big| \}.
\end{align}
Due to triangle inequality, the difference, in term of trace distance, between the operator in Eqn.~\ref{26} and  $\rho = H_i/\gamma_i$ is accumulated as:
\begin{align}
     \sum_{k_1,k_2,...,k_M=1}^d   \Big( \Tr \Big|  \ket{\lambda_{i_1} (k_1)}  \bra{\lambda_{i_1} (k_1)} - \sum_{i=1}^L p_j w_j(i_1,k_1) w_j(i_1,k_1)^\dagger \Big| + \cdots \\ +\Tr \Big| \ket{\lambda_{i_M} (k_M)}  \bra{\lambda_{i_M} (k_M)} - \sum_{j=1}^L p_j w_j(i_M,k_M) w_j(i_M,k_M)^\dagger \Big|   \Big),
\end{align}
which can be shown to be less than 
$$ \frac{1}{\gamma_i}  \sum_{\substack{k_1,k_2,...,k_M = 1 }}^d  \lambda_{i_1}(k_1) \lambda_{i_2}(k_2) \cdots \lambda_{i_M} (k_M) \Delta  = \Delta.$$
We note that this is the trace distance, which implies that in the operator norm-distance, the error is $\Delta $. Because the error of each (block-encoded) $\frac{H_i}{\gamma_i}$ is $\Delta$, the error accumulated for $H$ is still $\Delta$. The eventual error (quantified in operator norm) of the evolution operator is $\mathcal{O}( \sqrt{\Delta}\log \frac{1}{\epsilon})$ because as the error for block-encoding $H$ is $\Delta$ and an application of Lemma~\ref{lemma: qsvt} (with a polynomial of degree $\mathcal{O}(\log \frac{1}{\epsilon})$ results in the final error $\mathcal{O}( \sqrt{\Delta}\log \frac{1}{\epsilon})$ in the block-encoded $P(H)$ (where $P(.)$ is the Jacobi-Anger polynomial) which approximates $\exp(-iHt)$ with an additive error of $\epsilon$. So, the total error accumulated at this point is $ \mathcal{O}\Big( \sqrt{\Delta} \log (\frac{1}{\epsilon}) + \epsilon \Big)$. If we desire the overall error to be $\mathcal{O}(\delta)$ then we need to set $ \sqrt{\Delta} \log (\frac{1}{\epsilon}) + \epsilon  = \delta$. For simplicity, we choose $\epsilon = \sqrt{\Delta}$, which then implies that $\delta = \mathcal{O}(\sqrt{\Delta})$. The value of $\Delta$ depends critically on the optimization procedure underlying Lemma~\ref{lemma: randomapproximation} and therefore, in general, depends on the input states as well as the sparsity value $s$ that we chose. To our knowledge, there is no explicit analytical formula for the relation between the error versus sparsity parameter in Lemma \ref{lemma: randomapproximation}. So, the complexity for this approach for simulating $H$ up to an additive error $\mathcal{O}(\delta) = \mathcal{O}(\sqrt{\Delta})$ is 
$$ \mathcal{O}\Big( t  \sum_{i=1}^K \gamma_i \log \big ( L^M \sum_{i=1}^K r(H_i) s^M\big) \log \frac{1}{\sqrt{\Delta}} \Big). $$

Again, the complexity above can be reduced in the special cases as we discussed earlier (in Section \ref{sec: simplification}), where for each $i$, we then effectively deal with the new matrix $H_i' = \otimes_{ H_{i_j} \in R_i} H_{i_j}$. So, the complexity analysis is the same, except that the value of $M$ now replaced by $|R_i|$ and the value of $ r(H_i)$ is replaced by $r(H_i') = \prod_{H_{i_j} \in R_i} r (H_{i_j}) $, which results in the complexity for quantum part to be:
$$  \mathcal{O}\Big( t  \sum_{i=1}^K \gamma_i' \log \big ( L^{|R_i|} s^{|R_i|}\sum_{i=1}^K \prod_{H_{i_j} \in R_i} r (H_{i_j})  \big)  \log \frac{1}{\sqrt{\Delta}} \Big).$$
With the upper bound of $\gamma_i' = \mathcal{O}( ||H_i||_o d^{|R_i|})$ and $\prod_{H_{i_j} \in R_i} r (H_{i_j}) = \mathcal{O}(d^{|R_i|})$, the above can be simplified as
$$ \mathcal{O}\Big( t  K^2 d^{|R|} \log \big ( L^{|R|} s^{|R|} d^{|R|} \big)  \log \frac{1}{\sqrt{\Delta}} \Big)  = \mathcal{O}\Big( t  K^2 d^{|R|} |R| \log (Lsd) \log \frac{1}{\sqrt{\Delta}} \Big). $$
The complexity for the classical part is 
$$\mathcal{O}\Big( \sum_{i=1}^K s^{|R_i|} \big(\sum_{i=1}^K |R_i| \prod_{H_{i_j} \in R_i} r (H_{i_j}) \big)  \rm poly (d) \Big)  = \mathcal{O}\Big( K^2 s^{|R|}  |R| d^{|R|} \rm poly(d) \Big). $$
The total number of qubits is $  \mathcal{O}\left(  \sum_{i=1}^K L^{|R_i|} s^{|R_i|}  r(H_i)   \right)= \mathcal{O}\Big(  K L^{|R|} s^{|R|} d^{|R|}\Big)$.

\subsection{Analysis of time-dependent simulation algorithm Algo.~\ref{algo: timedependentsimulation}}
As described in the Algo.~\ref{algo: timedependentsimulation}, from Step 1 through Step 4, the procedure is the same as in the time-independent case. Therefore, the circuit complexity for this part is the same, which is $\mathcal{O}\Big( r(H_{i_1}) r(H_{i_2}) \cdots r(H_{i_M}) (M \log d)  \Big)$ if using Lemma~\ref{lemma: improveddme} (as described in Sec.~\ref{sec: alternativeapproach}). In the second step of Algo.~\ref{algo: timedependentsimulation} (which is technically a 5th step, as the first four steps are borrowed from the time-independent algorithm), we use a single-qubit rotation gate, so the gate complexity (of this step only) is $\mathcal{O}(1)$. However, we point out a subtlety that in reality, such a rotation gate is usually implemented by tuning a simple time-dependent Hamiltonian for some time, which is $t = \cos \frac{\theta}{2}$ in this case. So, in reality, the complexity of this step is $\mathcal{O}\big( t \big)$, which implies that the total complexity of the gate at this point is $ \mathcal{O}\Big( t \ r(H_{i_1}) r(H_{i_2}) \cdots r(H_{i_M}) (M \log d)  \Big) $. In the third step, we need to employ Lemma~\ref{lemma: qsvt} with an appropriate function  $P$ to obtain the block-encoding of $\beta_i(t)$, which incurs a quantum circuit of complexity $\mathcal{O}\big( \deg \beta_i(t) \big)$ (where $\deg (.)$ refers to the degree of the corresponding function).  In the fourth step, we need to use the lemma \ref{lemma: tensorproduct} to construct the block-encoding of $\beta_i(t) \frac{H_i}{\gamma_i}$, which uses the block-encoding of $\beta_i(t)$ and $\frac{H_i}{\gamma_i}$ one time each, so that the circuit complexity remains the same as in the previous step. The fifth step uses $K$ different block-encodings of the previous step, so the total complexity in this step is 
$$  \mathcal{O}\Big(  t \sum_{i=1}^K r(H_i) (M \log d)  \Big).$$
At the final steps, we first need to use Lemma~\ref{lemma: amp_amp} to remove the factor $\sum_{i=1}^K \gamma_i$ (as in Step 6 of Algo.~\ref{algo: timeindependentsimulation}) and then use Lemma~\ref{lemma: qsvt} with the Jacobi-Anger expansion that approximates $\exp(-ix)$ (we choose $t'=1$ in Step 7 of Algo.~\ref{algo: timedependentsimulation}). So, the total complexity for simulating $H(t)$ with a precision $\delta$ is
$$  \  \mathcal{O}\Big( \sum_{i=1}^K \gamma_i   \sum_{i=1}^K r(H_i)  t \  (M \log d)  \log \frac{1}{\delta}\Big).  $$
Finally, we note that the above complexity is for the quantum part. The classical part still has complexity $\mathcal{O}(d^3)$ as the time-independent case. In special cases as in Section~\ref{sec: simplification}, then the whole analysis above holds, except replacing $M$ by $|R_i|$ and $ r(H_i)= r(H_{i_1}) r(H_{i_2}) \cdots r(H_{i_M})$ is replaced by $r(H_i') = \prod_{H_{i_j} \in R_i} r (H_{i_j}) $, so the complexity is 
$$  \  \mathcal{O}\Big( \sum_{i=1}^K \gamma_i'   \big( \sum_{i=1}^K \prod_{H_{i_j} \in R_i} r (H_{i_j})  \big) t \  (|R_i| \log d)  \log \frac{1}{\delta}\Big). $$
Replacing the upper bound of $\gamma_i' = \mathcal{O}(d^{|R_i|})$ and $ \prod_{H_{i_j} \in R_i} r (H_{i_j}) = \mathcal{O}(d^{|R_i|})$ to the above, we arrive at the complexity
 $$  \mathcal{O}\Big( K^2 d^{2|R|}  t \ |R| \log (d)  \log \frac{1}{\delta}\Big).$$

\section{Discussion }
\label{sec: discussion}
For convenience in analyzing the efficient regime, we first summarize the above results in the Table \ref{tab: summarizingresults}. \\
\begin{table}[htbp]
    \centering
    \begin{tabular}{|c|c|c|}
    \hline
       & Approach 1. Sec.~\ref{sec: quantum} &  Approach 2. Sec.~\ref{sec: alternativeapproach} \\
        \hline
      Complexity (classical + quantum) & $ \mathcal{O}\Big( |R| K d^3 +  K^2 d^{|R|} (|R| \log d)     t \log \frac{1}{\delta}$    &  $ \mathcal{O}\Big( |R| K d^3 +  K^2  (|R| \log d)   \frac{d^{4|R|}}{\delta^2} t^3 \log \frac{1}{\delta}  \Big) $  \\
         \hline
     Total qubits  &   $\mathcal{O}\big(  M d\big)$ & $\mathcal{O} \big( M d \big)$\\
         \hline
    \end{tabular}
    \centering
    
    \begin{tabular}{|c|c|c|}
    \hline
   & Approach 3. Sec.~\ref{sec: thirdapproach} &  Time-dependent, commuting case. Sec.~\ref{sec: generalization}   \\
    \hline
   Complexity (classical + quantum)  & $\mathcal{O}\Big( K^2 s^{|R|}  |R| d^{|R|} \rm poly(d)  $   &  $\mathcal{O}(|R|Kd^3) +$ \\
    & $+ t  K^2 d^{|R|} |R| \log (Lsd) \log \frac{1}{\sqrt{\Delta}}  \Big)$  & $ \mathcal{O}\Big( K^2 d^{2|R|}  t (|R| \log d)   \log \frac{1}{\delta}\Big) $ \\
     \hline
     Total qubits & $  \mathcal{O}\left(   K L^{|R|} s^{|R|} d^{|R|} \right)$  & $\mathcal{O}\big( M d\big)$ \\
     \hline
    \end{tabular}
    \caption{Table summarizing the complexity of our hybrid classical-quantum algorithm for simulating Hamiltonian to an additive precision $\delta$. We remind that the value $|R|$ is defined as the maximum among $\{ |R_i|\}_{i=1}^K$. }
    \label{tab: summarizingresults}
\end{table}

\begin{table}[]
    \centering
    \begin{tabular}{|c|c|}
    \hline
      Existing works   &  Gate complexity\\
      \hline
      Ref.~\cite{berry2007efficient,berry2012black,berry2015hamiltonian, low2019hamiltonian}  (sparse-access model)   &  $\mathcal{\tilde{O}}\big( ||H||_{\max} t n s \log \frac{1}{\delta}\big)$  \\
      \hline
      Ref.~\cite{low2017optimal} (sparse-access model) & $\mathcal{O}\big(||H||_{\max}  t n s \log \frac{1}{\delta}\big) $ \\
      \hline
      Ref.~\cite{berry2015simulating} (LCU) & $\mathcal{O}\left( Mt (n + \log M) \log \frac{t}{\delta} \right) $ \\
      \hline
      Ref.~\cite{childs2019nearly, tran2020destructive} (lattice system) & $\mathcal{O}\big(  \frac{1}{\epsilon} (nt)^{1+\mathcal{O}(1)}  \big)$\\
      \hline
      Ref.~\cite{berry2020time} (sparse-access model) & $\mathcal{\tilde{O}}\big( \frac{1}{\delta} n s^4 t ( \int_0^1 d\tau ||H(\tau)||_{\max} )^2 \big)$ \\
      \hline
      Ref.~\cite{watkins2024time} (LCU) & $\mathcal{\tilde{O}}\big(  ||\alpha||_{\infty, 1}^{rev} t + \log(\frac{1}{\delta}) + t^2 \frac{1}{\delta}\max_{t} || \frac{dH}{dt}||_o  \big)$ \\
      \hline
      Ref.~\cite{kieferova2019simulating} (sparse-access model) & $\mathcal{\tilde{O}}\big( s^2 ||H||_{\max} t n \big)$ \\
      \hline
    \end{tabular}
    \caption{Table summarizing the complexities of existing quantum simulation algorithms with the corresponding input Hamiltonian model. In these works, the Hamiltonian $H$ is defined on $n$-qubits system and $s$ is the sparsity of $H$. $||H||_{\max}$ refers to the maximum element of $H$, $\delta$ is the error tolerance. In the Ref.~\cite{watkins2024time}, $||\alpha||_{\infty,1}^{rev}$ is defined as $ \sum_{j} \sup_t | \alpha_j(t)|$ where $\alpha_j$ refers to the $j$-th entry of the vector $\alpha$.   }
    \label{tab: existingwork}
\end{table}

\subsection{Analyzing the (most) efficient regime and relative comparison to existing literature}

\noindent
\textbf{Efficient regime. } First, we discuss in which regime our algorithm achieves the best performance. All approaches above have both the classical and quantum complexity that depend exponentially on $|R|$, which is defined as the maximum value among $ \{ |R_i|\}_{i=1}^K$. As such, these methods are most effective when $|R| = \mathcal{O}(1)$. In reality, there are many instances from quantum many-body systems that admit the structure where each $H_i= H_{i_1} \otimes H_{i_2} \otimes\cdots \otimes H_{i_M} $ and many of these smaller matrices are identity, e.g., lattice systems with nearest-neighbor interaction terms. As such, we believe that our method is most effective when dealing with these kinds of systems. \\

\noindent
\textbf{Brief summary of existing works. } There has been a rich literature in the field of quantum simulation. There are mainly two input models/assumptions that are present, including the sparse-access model and the linear combination of unitaries model (LCU). In the first model, for example, as in~\cite{berry2007efficient,berry2012black, low2017optimal,low2019hamiltonian, aharonov2003adiabatic}, it is assumed that there are two oracles. The first can efficiently query the location of the entries in $H$, and the second can efficiently query the value of such an entry in $H$. The most efficient (and even simplest) algorithm for simulating Hamiltonian under this model is the so-called quantum signal processing/QSVT~\cite{low2017optimal,low2019hamiltonian}, achieving the optimal complexity (in most parameters), i.e., $\mathcal{\tilde{O}}\big(t  n s  \log \frac{1}{\delta}\big)$, where $s$  is the sparsity of $H$, $n$ is the number of qubits, $\delta$ is the precision, and $\mathcal{\tilde{O}}$  hides the polylogarithmic terms. At the same time, the work of~\cite{berry2015hamiltonian} considers the linear combination of unitaries (LCU) model, in which the Hamiltonian has the form $H = \sum_{i=1}^M \alpha_i U_i$. Provided that each $U_i$ can be efficiently implemented, their algorithm also achieves (nearly) optimal scaling in all parameters, i.e., $ \mathcal{O}\left( Mt (n + \log M) \log \frac{t}{\epsilon} \right) $. Some other notable works, including lattice simulations~\cite{tran2020destructive, childs2019nearly}, assume the lattice Hamiltonian $H = \sum_i H_i$, similar to ours. However, their strategy relies on decomposing the Hamiltonian $H$ into several parts, each containing commuting pairwise terms,  and applying the product formula appropriately. It results in a lattice simulation algorithm with a (nearly) optimal complexity $\mathcal{O}\left( n^2 \frac{t}{\epsilon}  \right)$. The work \cite{chen2021quantum} considers the time-dependent Hamiltonian of the form $H(t) = \sum_{i=1}^M D_i(t) P_i$ where $D_i(t)$ is a diagonal operator with time-dependent entries and $P_i$ is the permutation operator. Their algorithm's strategy relies on the permutation expansion and achieves almost optimal complexity. 

Although there are many more quantum simulation algorithms in addition to the aforementioned ones, especially in the time-dependent setting~\cite{berry2020time, watkins2024time, an2022time, kieferova2019simulating, cao2025unifying} (see Table~\ref{tab: existingwork}), their input assumptions are fundamentally different from ours, and thus it might not be most appropriate to compare theirs to ours on an equal footing. Moreover, most existing works (except~\cite{berry2015hamiltonian}) rely on the oracle assumption (sparse-access model). Whether this oracle can be efficiently implemented for any Hamiltonian remains unclear. The complexity presented in those works involving the sparse-access model does not include the complexity of realizing such an oracle. On the other hand, the LCU model does not explicitly require the oracle, but it comes with an assumption that all unitaries $\{U_i\}$ can be efficiently realized. Furthermore, their approach is only accessible to a certain type of Hamiltonian, and thus is not general overall. \\ 

\noindent
\textbf{Complementary to existing work.} As we have just discussed, the input assumptions of our work and the existing ones are different. Therefore, we believe that our work can \textit{complement them very well and can work when the same input assumptions from these other works cannot be met}. Our algorithm is aimed at a time-independent Hamiltonian, but it could be adapted to a certain type of time-dependent Hamiltonian. In addition, our work does not require the oracle as in the sparse-access model, and our algorithm can generally perform well even when $H$ is not sparse. In the case where the sparsity is as large as $2^n$ (equal to the dimension of the Hamiltonian), our algorithm can even achieve an exponential speedup in the number of qubits. In a relative comparison to the LCU model in~\cite{berry2015hamiltonian}, our algorithm can deal with a more general type of Hamiltonians where each $H_i$ is not necessarily a unitary, and we only need to have classical information about the tensor structure of each $H_i$. At the same time, our algorithm can also deal with lattice simulation as in Ref.~\cite{childs2019nearly, tran2020destructive}. In this setting, the Hamiltonian for the lattice system has exactly the same form as what we consider in this work. If we use the method of Sec.~\ref{sec: alternativeapproach} or Sec.~\ref{sec: thirdapproach}, then our algorithm for lattice simulation achieves logarithmic scaling on $\frac{1}{\delta}$, thus exhibiting exponential speedup compared to Ref.~\cite{tran2020destructive, childs2019nearly}. In addition, we note that our method can also handle certain time-dependent lattice systems, thus complementing prior works~\cite{childs2019nearly, tran2020destructive}.

\subsection{Corollaries: application of \cite{harrow2025randomized} to quantum simulation and quantum state preparation}
We recall from Section~\ref{sec: thirdapproach} that we have used the result of~\cite{harrow2025randomized} (Lemma~\ref{lemma: randomapproximation}) to find a good approximation to the eigenstates of each $H_{i_j}$. These approximations are then employed within a quantum procedure (with Lemma~\ref{lemma: statepreparation} and Lemma~\ref{lemma: improveddme}) to obtain the block-encoding of the Hamiltonian $H$ of interest, from which the evolution operator can be obtained. As a consequence, we have provided a way to leverage the result of~\cite{harrow2025randomized} toward quantum simulation, which was stated as an open question in the same work. 

Here, we point out that this result is also applicable to the problem of quantum state preparation, or more precisely, it can be used to practically enhance the state preparation protocol proposed in~\cite{zhang2022quantum}. The problem of quantum state preparation is simply stated as follows: let $\ket{\Phi} = \sum_{i=1}^{2^n} a_i \ket{i-1}$ be an $n$-qubit quantum state (apparently with the normalization $\sum_{i=1}^{2^n} |a_i|^2 = 1$) of interest; the goal is to construct an efficient quantum circuit $U$ (of complexity $\mathcal{O}\big( n  \big)$) so that $U \ket{0}^{\otimes n} = \ket{\Phi}$. A more relaxed version of this problem allows us to use ancilla qubits, and the state $\ket{\Phi}$ might be obtained upon an appropriate measurement of the ancilla qubits. There are many results in this direction, including~\cite{grover2002creating, grover2000synthesis, zhang2022quantum, nakaji2022approximate, zoufal2019quantum, zhang2021low, mcardle2022quantum, plesch2011quantum, marin2023quantum}, with many types of techniques introduced. For example, Ref.~\cite{nakaji2022approximate} relies on the variational strategy, while Ref.~\cite{mcardle2022quantum} is built on the quantum singular value transformation framework. The work~\cite{grover2002creating, grover2000synthesis, marin2023quantum}, while efficient, can only be applied to a state with efficiently integrable amplitudes. The most general and universal state preparation protocol is given in Ref.~\cite{zhang2022quantum}. By general and universal, we meant that their method can be used to prepare a quantum state with any structure. For an $n$-qubit quantum state $\ket{\Phi}$ having $s$ (for $s \leq 2^n$) nonzero entries, their method uses $\mathcal{O}(s)$ ancilla qubits and a quantum circuit of depth $\mathcal{O}\big(  \log (sn)  \big)$. Although the depth is efficient in the dimension of the state $\ket{\Phi}$, this method is practically only efficient when the sparsity $s$ is not large, e.g., of order $\mathcal{O}\big( \rm poly (n) \big)$. Otherwise, there will be an exponential number of qubits required, which implies inefficiency. 

In the following, we will describe how to leverage the result of Ref.~\cite{harrow2025randomized} to make the state preparation protocol in Ref.~\cite{zhang2022quantum} more efficient in practice. First, from the classical knowledge of the amplitudes of the state $\ket{\Phi} = \sum_{i=1}^{2^n} a_i \ket{i-1} $, we use Lemma~\ref{lemma: randomapproximation} to find the probabilities $\{p_i\}_{i=1}^L$ and states $\{ w_i\}_{i=1}^L$ that form the ``good'' approximation to the state $\ket{\Phi}$ of interest. As indicated in Lemma~\ref{lemma: randomapproximation}, the sparsity of those states $\{ w_i\}_{i=1}^L$ can be chosen (and the value of $L$ does not depend on the input dimension $2^n$), so we choose the sparsity $S$ so that $S$ is significantly smaller than the dimension $2^n$. 

To this end, we show in Appendix~\ref{sec: errorbound} that, in the context of Lemma~\ref{lemma: randomapproximation}, if $\Tr \Big|  v v^\dagger - \sum_{j=1}^L p_j w_j  w_j^\dagger \Big| \leq \epsilon 
$, then $\Big|  v- \sum_{j=1}^L \sqrt{p_j}w_j \Big|_2 \leq \sqrt{\epsilon}$. This means that $\sum_{j=1}^L \sqrt{p_j}w_j$ is a good approximation to $v$, so our goal now is to prepare $\sum_{j=1}^L \sqrt{p_j}w_j$. As each $w_j$ is a quantum state of sparsity $s$, we can use Lemma~\ref{lemma: statepreparation} to obtain a unitary, denoted by $U_j$ that prepares $w_j$, i.e., $U_j \ket{0}^n = w_j$. Next, we use Lemma~\ref{lemma: sumencoding} to construct the block-encoding of $ \frac{1}{\sum_{i=1}^L \sqrt{p_j}}\sum_{i=1}^L \sqrt{p_j}U_j$. Then we take this unitary block-encoding to apply to the state $\ket{\bf 0} \ket{0}^n$ (where $\ket{\bf 0}$ accounts for the ancilla qubits required for block-encoding purposes), then according to the definition~\ref{def: blockencode}, we obtain the state:
\begin{align}
    \ket{\bf 0}  \frac{1}{\sum_{i=1}^L \sqrt{p_j}} \sum_{i=1}^L \sqrt{p_j}U_j \ket{0}^n + \ket{\rm Garbage} =  \frac{1}{\sum_{i=1}^L \sqrt{p_j}}\ket{\bf 0}  \sum_{i=1}^L \sqrt{p_j} w_j    + \ket{\rm Garbage}.
\end{align}
Measuring the ancilla qubits and post-select $\ket{\bf 0}$, then we obtain the desired state $\sum_{i=1}^L \sqrt{p_j}w_j$ with the success probability $\big|  \frac{1}{\sum_{i=1}^L \sqrt{p_j}}\big|^2 $. 

To analyze the overall complexity, each $w_j$ has a sparsity $S$ (which can be chosen to be much smaller than the dimension $2^n$), so the circuit depth of each $U_j$ is $\mathcal{O}\Big(\log (Sn)\Big)$, with an additional $\mathcal{O}(s)$ ancillary qubits. The next step is to use Lemma~\ref{lemma: sumencoding}, which uses each $U_j$ one time, thus incurring a total quantum circuit depth $\mathcal{O}\Big( L  \log (Sn)  \Big)$, with a total of $\mathcal{O}(S)$ ancilla qubits. Given that $S \ll 2^n$, this complexity is much more efficient in terms of circuit depth, particularly when the desired state $\ket{\Phi}$ is dense. Recall that the trade-off we have is that the outcome is not an exact state but rather an approximation to the target state. Finally, in the context of Corollary~\ref{corollary: statepreparation}, the value of $\epsilon$ is different from the discussion we have had in this section only by a slight abuse of notation, i.e., $\sqrt{\epsilon}$ we used above is $\epsilon$ we used in Corollary~\ref{corollary: statepreparation}.

\section{Conclusion}
\label{sec: conclusion}
In this work, we have introduced a hybrid framework for simulating  Hamiltonians of a certain form. Given the classical information of the matrices that make up the main Hamiltonian, we use the classical algorithm to diagonalize them. The output from this diagonalization step is then used as input to a quantum procedure, which converts it into the block encoding of the main Hamiltonian. The evolution operator is then obtained using standard techniques from block encoding/quantum singular-value transformation. In particular, we have discussed three approaches for obtaining the block encoding of the Hamiltonian of interest. Each approach has certain strengths and shortcomings. In addition, our hybrid framework can be extended to the time-dependent Hamiltonian, in which the terms $\{H_i\}_{i=1}^K$ pairwise commute. Particularly, as a byproduct of our method, we have shown how to leverage the work of~\cite{harrow2025randomized} to enhance the state preparation protocol~\cite{zhang2022quantum} in practice. 

Overall, the hybrid route we take in this work provides an alternative approach to the problem of Hamiltonian simulation, which is regarded as one of the most important applications of quantum computers. We have particularly shown that from a broader perspective, our hybrid framework can act as a complement to the existing quantum simulation algorithms. The input assumptions or access to the main Hamiltonian in the existing quantum simulation context might not be friendly to near-term devices (probably except~\cite{berry2015simulating}). At the same time, it can be met quite simply in our case, as in reality,  the tensor structure as well as the concrete composing matrices are classically known (or can be classically derived). Thus, we believe that our hybrid framework can take advantage of both classical and (near-term) quantum resources, thus extending the reach to include more types of Hamiltonians that could be simulated. That being said, our method can only deal with the time-dependent Hamiltonian in which the composing terms commute, which is quite restricted. In the above, we have pointed out a way, inspired by Lie-Trotter-Suzuki formula, to handle the general case. However, we believe that this approach will incur a large amount of error, which can incur a large dependence on inverse of error tolerance. How to extend the idea and method in this work to handle more general time-dependent Hamiltonian is thus left for future research.

\section*{Acknowledgments}
This work was supported by the U.S. Department of Energy, Office of Science, National Quantum Information Science Research Centers, Co-design Center for Quantum Advantage (C2QA) under Contract No. DE-SC0012704. The authors also acknowledge support from the Center for Distributed Quantum Processing at Stony Brook University. 

\bibliography{ref.bib}
\bibliographystyle{unsrt}

\appendix
\onecolumngrid

\section{Block-encoding and quantum singular value transformation}
\label{sec: summaryofnecessarytechniques}
We briefly summarize the essential quantum tools used in our algorithm. For conciseness, we highlight only the main results and omit technical details, which are thoroughly covered in~\cite{gilyen2019quantum}. An identical summary is also presented in~\cite{lee2025new}.

\begin{definition}[Block-encoding unitary, see e.g.~\cite{low2017optimal, low2019hamiltonian, gilyen2019quantum}]
\label{def: blockencode} 
Let $A$ be a Hermitian matrix of size $N \times N$ with operator norm $\norm{A} < 1$. A unitary matrix $U$ is said to be an \emph{exact block encoding} of $A$ if
\begin{align}
    U = \begin{pmatrix}
       A & * \\
       * & * \\
    \end{pmatrix},
\end{align}
where the top-left block of $U$ corresponds to $A$. Equivalently, one can write
\begin{equation}
    U = \ket{\mathbf{0}}\bra{\mathbf{0}} \otimes A + (\cdots),    
\end{equation}
where $\ket{\mathbf{0}}$ denotes an ancillary state used for block encoding, and $(\cdots)$ represents the remaining components orthogonal to $\ket{\mathbf{0}}\bra{\mathbf{0}} \otimes A$. If instead $U$ satisfies
\begin{equation}
    U = \ket{\mathbf{0}}\bra{\mathbf{0}} \otimes \tilde{A} + (\cdots),
\end{equation}
for some $\tilde{A}$ such that $\|\tilde{A} - A\| \leq \epsilon$, then $U$ is called an {$\epsilon$-approximate block encoding} of $A$. Furthermore, the action of $U$ on a state $\ket{\mathbf{0}}\ket{\phi}$ is given by
\begin{align}
    \label{eqn: action}
    U \ket{\mathbf{0}}\ket{\phi} = \ket{\mathbf{0}} A\ket{\phi} + \ket{\mathrm{Garbage}},
\end{align}
where $\ket{\mathrm{Garbage}}$ is a state orthogonal to $\ket{\mathbf{0}}A\ket{\phi}$. The circuit complexity (e.g., depth) of $U$ is referred to as the {complexity of block encoding $A$}.
\end{definition}

\begin{lemma}[Amplification, Theorem 30 of~\cite{gilyen2019quantum}]
\label{lemma: amp_amp}
Let $U$, $\Pi$, $\widetilde{\Pi} \in {\rm End}(\mathcal{H}_U)$ be linear operators on $\mathcal{H}_U$ such that $U$ is a unitary, and $\Pi$, $\widetilde{\Pi}$ are orthogonal projectors. 
Let $\gamma>1$ and $\delta,\epsilon \in (0,\frac{1}{2})$. 
Suppose that $\widetilde{\Pi}U\Pi=W \Sigma V^\dagger=\sum_{i}\varsigma_i\ket{w_i}\bra{v_i}$ is a singular value decomposition. 
Then there is an $m= \mathcal{O} \Big(\frac{\gamma}{\delta}
\log \left(\frac{\gamma}{\epsilon} \right)\Big)$ and an efficiently computable $\Phi\in\mathbb{R}^m$ such that
\begin{align}
\left(\bra{+}\otimes\widetilde{\Pi}_{\leq\frac{1-\delta}{\gamma}}\right)U_\Phi \left(\ket{+}\otimes\Pi_{\leq\frac{1-\delta}{\gamma}}\right) = \sum_{i\colon\varsigma_i\leq \frac{1- \delta}{\gamma} }\tilde{\varsigma}_i\ket{w_i}\bra{v_i} , \text{ where } \Big|\!\Big|\frac{\tilde{\varsigma}_i}{\gamma\varsigma_i}-1 \Big|\!\Big|\leq \epsilon.
\end{align}
Moreover, $U_\Phi$ can be implemented using a single ancilla qubit with $m$ uses of $U$ and $U^\dagger$, $m$ uses of C$_\Pi$NOT and $m$ uses of C$_{\widetilde{\Pi}}$NOT gates and $m$ single qubit gates.
Here,
\begin{itemize}
\item C$_\Pi$NOT$:=X \otimes \Pi + I \otimes (I - \Pi)$ and a similar definition for C$_{\widetilde{\Pi}}$NOT; see Definition 2 in \cite{gilyen2019quantum},
\item $U_\Phi$: alternating phase modulation sequence; see Definition 15 in \cite{gilyen2019quantum},
\item $\Pi_{\leq \delta}$, $\widetilde{\Pi}_{\leq \delta}$: singular value threshold projectors; see Definition 24 in \cite{gilyen2019quantum}.
\end{itemize}
\end{lemma}
Based on~\ref{def: blockencode}, several properties, though immediate, are of particular importance and are listed below.
\begin{remark}[Properties of block-encoding unitary]
The block-encoding framework has the following immediate consequences:
\begin{enumerate}[label=(\roman*)]
    \item Any unitary $U$ is trivially an exact block encoding of itself.
    \item If $U$ is a block encoding of $A$, then so is $\Ibb_m \otimes U$ for any $m \geq 1$.
    \item The identity matrix $\Ibb_m$ can be trivially block encoded, for example, by $\sigma_z \otimes \Ibb_m$.
\end{enumerate}
\end{remark}

Given a set of block-encoded operators, various arithmetic operations can be done with them. Here, we simply introduce some key operations that are especially relevant to our algorithm, focusing on how they are implemented and their time complexity, without going into proofs. For more detailed explanations, see~\cite{gilyen2019quantum, camps2020approximate}.

\begin{lemma}[Informal, product of block-encoded operators, see e.g.~\cite{gilyen2019quantum}]
\label{lemma: product}
    Given unitary block encodings of two matrices $A_1$ and $A_2$, with respective implementation complexities $T_1$ and $T_2$, there exists an efficient procedure for constructing a unitary block encoding of the product $A_1 A_2$ with complexity $T_1 + T_2$.
\end{lemma}

\begin{lemma}[Informal, tensor product of block-encoded operators, see e.g.~{\cite[Theorem 1]{camps2020approximate}}]\label{lemma: tensorproduct}
    Given unitary block-encodings $\{U_i\}_{i=1}^m$ of multiple operators $\{M_i\}_{i=1}^m$ (assumed to be exact), there exists a procedure that constructs a unitary block-encoding of $\bigotimes_{i=1}^m M_i$ using a single application of each $U_i$ and $\mathcal{O}(1)$ SWAP gates.
\end{lemma}

\begin{lemma}[Informal, linear combination of block-encoded operators, see e.g.~{\cite[Theorem 52]{gilyen2019quantum}}]
    We are given the unitary block encoding of multiple operators $\{A_i\}_{i=1}^m$ and $\{\alpha_i\}_{i=1}^m$ satisfying $\sum_{i=1}^m |\alpha_i | = 1$. Then, given a unitary that prepare the states  $ \sum_{i=1}^m \sqrt{|\alpha_i|} \ket{i}$,  there is a procedure that produces a unitary block encoding operator of $\sum_{i=1}^m \alpha_i A_i $ in time complexity $\mathcal{O}(m)$, e.g., using the block encoding of each operator $A_i$ a single time. 
    \label{lemma: sumencoding}
\end{lemma}

\begin{lemma}[Informal, scaling multiplication of block-encoded operators] 
\label{lemma: scale}
    Given a block encoding of some matrix $A$, as in~\ref{def: blockencode}, the block encoding of $A/p$ where $p > 1$ can be prepared with an extra $\mathcal{O}(1)$ cost.
\end{lemma}



\begin{lemma}[Matrix inversion, see e.g.,~\cite{gilyen2019quantum, childs2017quantum}]\label{lemma: matrixinversion}
Given a block encoding of some matrix $A$  with operator norm $||A|| \leq 1$ and block-encoding complexity $T_A$, then there is a quantum circuit producing an $\epsilon$-approximated block encoding of ${A^{-1}}/{\kappa}$ where $\kappa$ is the conditional number of $A$. The complexity of this quantum circuit is $\mathcal{O}\left( \kappa T_A \log \left({1}/{\epsilon}\right)\right)$. 
\end{lemma}


\section{Proof that $\Big| v- \sum_{i=1}^L \sqrt{p_j}w_j \Big|_2 \leq \sqrt{\epsilon}$}
\label{sec: errorbound}
In this section, we will show that if 
\begin{align}
     \Tr \Big|  v v^\dagger - \sum_{j=1}^L p_j w_j  w_j^\dagger \Big| \leq \epsilon 
\end{align}
Then $\sum_{i=1}^L \sqrt{p_j}w_j  $ is a good approximation to $v$. Similarly to the main text, we assume $v, w_j \in \mathbb{C}^d$ for all $j$. Denote the diagonal entries of $v$ as $v_1,v_2,...,v_d$, of $w_j$ as $w_{j 1}, w_{j2}, ..., w_{j d}$. Then the inequality above is:
\begin{align}
  \sum_{i=1}^d     \Big| |v_i|^2 - \sum_{j=1}^L p_j |w_{ji }|^2   \Big| \leq \epsilon
\end{align} 
First, we point out that, for $0 \leq x \leq 1 $ then $ x < \sqrt{x}$. So for all $j$, it holds that $ p_j |w_{ji}|^2 \leq \sqrt{p_j}|w_{ji}|$, which implies that $ -\sum_{j=1}^L p_j |w_{ji }|^2 > -\sum_{j=1}^L\sqrt{p_j}|w_{ji}|  $. It means that:
\begin{align}
  \sum_{i=1}^d     \Big| |v_i|^2 - \sum_{j=1}^L\sqrt{p_j}|w_{ji}|   \Big| \leq  \sum_{i=1}^d     \Big| |v_i|^2 - \sum_{j=1}^L p_j |w_{ji }|^2   \Big| \leq \epsilon
\end{align}
Furthermore, because $\sum_{j=1}^L\sqrt{p_j}|w_{ji}| \approx v_i $ which is less than 1, so we have that $\sum_{j=1}^L\sqrt{p_j}|w_{ji}| \geq (\sum_{j=1}^L\sqrt{p_j}|w_{ji}|)^2 $. Combining with the inequality above, we have:
\begin{align}
 \sum_{i=1}^d     \Big| |v_i|^2 - \big( \sum_{j=1}^L\sqrt{p_j}|w_{ji}| \big)^2  \Big|  \leq  \sum_{i=1}^d     \Big| |v_i|^2 - \sum_{j=1}^L\sqrt{p_j}|w_{ji}|   \Big| \leq \epsilon
\end{align}
Second, we point out the following inequality: for $a,b\in \mathbb{C}$, we have the following:
\begin{align}
   \Big| (|a| - |b|)^2 \Big|  &= \Big| (|a|-|b|) (|a|-|b|) \Big| \\
                            &\leq  \Big| (|a|-|b|) (|a| + |b|)  \Big|   \\
                            & =  \Big|  |a|^2  - |b|^2   \Big| 
\end{align}
So $\Big| |v_i|^2 - \big( \sum_{j=1}^L\sqrt{p_j}|w_{ji}| \big)^2  \Big| \geq \Big|  \big( |v_i| - \big( \sum_{j=1}^L\sqrt{p_j}|w_{ji}| \big)^2  \Big| $ for all $i$, which implies that: 
\begin{align}
    \sum_{i=1}^d \Big|  \big( |v_i| - \big( \sum_{j=1}^L\sqrt{p_j}|w_{ji}| \big)^2  \Big| \leq \sum_{i=1}^d \Big| |v_i|^2 - \big( \sum_{j=1}^L\sqrt{p_j}|w_{ji}| \big)^2  \Big| \leq \epsilon
\end{align}
The left-most term $ \sum_{i=1}^d \Big|  \big( |v_i| - \big( \sum_{j=1}^L\sqrt{p_j}|w_{ji}| \big)^2  \Big|  $ is exactly $ \Big| v- \sum_{i=1}^L \sqrt{p_j}w_j \Big|_2^2$, which means that $ \Big| v- \sum_{i=1}^L \sqrt{p_j}w_j \Big|_2 \leq \sqrt{\epsilon}  $. Therefore, $\sum_{i=1}^L \sqrt{p_j}w_j$ is a good approximation to $v$.

\end{document}